\documentclass[a4paper,11pt]{article}
\usepackage[utf8]{inputenc}
\usepackage[T1]{fontenc}

\usepackage{jheppub}
\usepackage{graphicx}
\usepackage{amsmath}
\usepackage{amsfonts}
\usepackage[utf8]{inputenc}
\usepackage[symbol]{footmisc}
\usepackage{braket}
\usepackage{xcolor}

\graphicspath{{./figures/}}

\usepackage{xspace}
\newcommand{\nubb}{\ensuremath{0\nu\beta\beta}\xspace}
\newcommand{\mbb}{\ensuremath{m_{\beta\beta}}\xspace}
\newcommand{\Ge}{\ensuremath{^{76}}\text{Ge}\xspace}
\newcommand{\Se}{\ensuremath{^{82}}\text{Se}\xspace}
\newcommand{\Mo}{\ensuremath{^{100}}\text{Mo}\xspace}
\newcommand{\Te}{\ensuremath{^{130}}\text{Te}\xspace}
\newcommand{\Xe}{\ensuremath{^{136}}\text{Xe}\xspace}

\title{Probing the Mechanism of Neutrinoless Double-Beta Decay in Multiple Isotopes}

\author{Matteo Agostini,}
\author{Frank F. Deppisch,}
\author{Graham Van Goffrier}
\date{\today}

\affiliation{Department of Physics and Astronomy, University College London, \\Gower Street, London WC1E 6BT, UK}

\emailAdd{matteo.agostini@ucl.ac.uk}
\emailAdd{f.deppisch@ucl.ac.uk}
\emailAdd{graham.vangoffrier.19@ucl.ac.uk}

\abstract{
A large experimental program is being mounted to search for neutrinoless double-beta decay over the next decade. Multiple experiments using different target isotopes are being prepared to explore the whole parameter space allowed for inverted-ordered light neutrinos, and have the potential to make discoveries in several other scenarios, including normal-ordered light neutrinos and other exotic mechanisms. We investigate to what extent long-range and exotic short-range contributions may be distinguished by combining measurements of the decay half-life across isotopes in the framework of a global Bayesian analysis. We demonstrate how measurements in two isotopes will constrain the parameter space up to a two-fold degeneracy, and how a further measurement in a third isotope removes such a degeneracy. We also discuss the impact of uncertainties and correlations in nuclear matrix element calculations. Our work motivates an experimental program measuring neutrinoless double-beta decay in more than one isotope, as this would break parameter degeneracies and advance our understanding of particle physics beyond the Standard Model.}

\begin{document}
\maketitle

\section{Introduction} 
\label{introduction}

The discovery that neutrinos are massive particles~\cite{Kajita:2016cak,McDonald:2016ixn} is, so far, our only laboratory evidence of physics beyond the Standard Model (SM). If neutrino masses are generated through the standard Higgs mechanism, then new particles must exist, i.e., the right-handed sterile neutrino counterparts. Alternatively, neutrino masses could be explained if neutrinos are their own antiparticles, i.e., they are Majorana fermions~\cite{Majorana:1937vz,Racah:1937qq}. These emerge in many New Physics scenarios where total lepton number, an accidental quantum number in the SM, is broken. If its associated symmetry breaking occurs at a very high energy scale or is sequestered from the SM sector, it can also explain why neutrinos are so much lighter than the other fermions~\cite{King2004}.  Such scenarios often give rise to a leptogenesis mechanism which can generate an asymmetry between the amount of matter and antimatter in the universe~\cite{Davidson2008}. Thus, the origin of matter and neutrino masses could be deeply coupled to neutrinos and New Physics beyond the SM.

The most sensitive and direct probe for Majorana neutrinos is the search for neutrinoless double-beta ($0\nu\beta\beta$) decay, a hypothetical nuclear process in which two neutrons simultaneously convert into two protons and two electrons~\cite{Agostini:2022zub}. In most models, the process predominantly takes place through the exchange of light, active Majorana neutrinos via SM left-handed currents, the so-called mass mechanism. However, New Physics scenarios that generate light neutrino masses typically will give rise to other contributions, mediated by exotic particles. Such contributions can be systematically described through effective operators \cite{doi:1981yb, Doi:1983, Tomoda:1990rs, Ali:2006iu, Ali:2007ec, Cirigliano:2017djv, Graf:2018ozy, Deppisch:2020ztt, Kotila:2021xgw}, capturing all scenarios at scales above the $0\nu\beta\beta$ decay momentum scale of $q \approx 100$~MeV.

While the discovery of $0\nu\beta\beta$ decay will prove that neutrinos have nonzero Majorana masses~\cite{Schechter:1981bd}, it will thus immediately evoke the question of what mechanism mediates the decay. There are many ways to resolve this, such as measuring the angular distribution of the $0\nu\beta\beta$ decay electrons \cite{Ali:2006iu, SuperNEMO:2010wnd}, probing $0\nu\beta\beta$ decay to excited states, or correlating the $0\nu\beta\beta$ decay measurement with other probes such cosmological observations \cite{Deppisch:2004kn}. Another possibility, explored and quantified in this article, is to measure the half-life of $0\nu\beta\beta$ decay in more than one isotope as first considered in \cite{Deppisch:2006hb}.

Over the last decades, a broad experimental program has been mounted to search for $0\nu\beta\beta$ decay. Several detection concepts have been developed and tested since then, leading to constraints on the decay half-life of the order of $10^{25}--10^{26}$~yr in several isotope~\cite{collaboration2019improved, GERDA:2020xhi, Gando:2012zm}. The field is now at a turning point. The scientific community and funding agencies worldwide are discussing the construction in the next decade of ton-scale experiments capable of reaching a sensitivity to half-lives values up to $10^{28}$~yr, fully exploring the parameter space allowed by assuming that neutrinos have Majorana masses following the so-called inverted ordering. Mature designs are already available for three experiments based on different isotopes: CUPID with \Mo~\cite{Pavan:2020ipz}, LEGEND with \Ge~\cite{Zsigmond:2020bfx}, and nEXO with \Xe~\cite{Pocar:2020zqz}.

In this work, we study the combined discovery power of these three experiments and their capability to pin down the mechanism mediating the decay by measuring its half-life in multiple isotopes. Nuclear physics plays a key role in this discussion, as it introduces uncertainties but also breaks degeneracies, ultimately enabling us to discriminate among potential $0\nu\beta\beta$ decay mechanisms. 

Previous research has argued that the ratio of light-neutrino-exchange to exotic nuclear matrix elements (NMEs) is strongly degenerate across the isotopes in which $0\nu\beta\beta$ decay occurs. Ref.~\cite{Faessler2011} demonstrated this effect for light-neutrino-exchange, heavy-neutrino-exchange, and two R-parity violating (RPV) mechanisms across the isotopes \Ge, \Se, \Mo and \Te. Meanwhile, Ref.~\cite{Lisi2015} produced a similar result for light-neutrino-exchange and heavy-neutrino-exchange across the isotopes \Ge, \Te, and \Xe.
If this degeneracy were too strong, it would pose a fundamental obstacle to our proposed analysis, as distinct contributions to $0\nu\beta\beta$ decay would not be separable even with observations in multiple isotopes. 

However, the situation has drastically changed in the last few years. Firstly, Ref.~\cite{Menendez18} has pointed out that nuclear-structure correlations, presently unaccounted for or only partially accounted for in all nuclear many-body methods, should contribute less to exotic mechanisms than to light-neutrino-exchange. In particular, such correlations appear to suppress NME contributions from internucleon distances $\geq 2$~fm, which are already near-irrelevant for the heavy-neutrino-exchange NME. Furthermore, exotic mechanisms with similar spin-structures to heavy-neutrino-exchange are expected to show a similar suppressed dependence on nucleon correlations. This new analysis does not contradict the findings of Ref.~\cite{Faessler2011} and \cite{Lisi2015}, and indeed explicitly supports them when their state spaces are restricted to uncorrelated nucleons. Secondly, estimates of five distinct exotic mechanisms~\cite{Deppisch:2020ztt} also demonstrated significant deviations from the isotopic degeneracy of the ratio between light-neutrino-exchange and exotic NMEs. The consequently enhanced exotic-mechanism NMEs for \Mo from Ref.~\cite{Deppisch:2020ztt} will play a key role in our analysis. Ref.~\cite{Graf:2022lhj} further explores the prospect of distinguishing 32 low-energy effective operators, both from half-life ratios and differing phase-space observables in multiple isotopes. Finally, uncertainties in light-neutrino-exchange NME estimation are known to partially stem from short-range nuclear physics which is systematic across isotopes, for example the problem of the quenching of the axial coupling $g_A$, well-known across $\beta$ decays \cite{Agostini:2022zub}, and contact terms to the two-nucleon potential identified within a chiral effective field theory framework by \cite{Cirigliano:2021qko}. As a consequences, positive statistical correlations occur between NME estimation biases in distinct nuclear isotopes, which should improve discrimination between mechanisms via half-life ratios \cite{Graf:2022lhj}.

In this paper, we quantify the extent to which observations of $0\nu\beta\beta$ decay in multiple isotopes may simultaneously constrain the parameter space for both light-neutrino-exchange and a single (i.e., RPV) exotic mechanism. This is guaranteed to be possible so long as the aforementioned degeneracy of the light-neutrino-exchange and exotic NME ratio is not exact. We propose a method to quantify uncertainties and correlation among current light-neutrino exchange NME estimations, and study the impact of future, potentially more accurate calculations. By developing tools and methodologies for a multi-mechanism analysis from multi-isotope observations, and thereby exploring the potential of such an analysis, we set the necessary groundwork for future analyses on real observations of $0\nu\beta\beta$ decay.

The paper is organized as follows. In Sec.~\ref{sec:mechanisms}, we discuss mechanisms that give rise to $0\nu\beta\beta$ decay and how they are combined to calculate the decay half-life. Sec.~\ref{sec:concept} describes the key concept of how multi-isotope measurements can pin down the mechanisms responsible for the decay. Sec.~\ref{sec:nmes} covers the nuclear physics and, specifically, how we estimate statistical uncertainties and correlations among NME calculations and Sec.~\ref{statistics} describes how the statistical experimental likelihood, combining hypothetical future measurements, is constructed. In Sec.~\ref{sec:results}, we present our sensitivity projections under two scenarios. In the first one, the decay is mediated by the exchange of Majorana neutrinos, and we quantify the strength of future constraints on other mechanisms. In the second scenario, the Majorana neutrino exchange is not the primary mechanism, and we evaluate the confidence with which we can exclude the absence of additional contributions. We conclude and provide an outlook in Sec.~\ref{conclusion}. Details on our statistical analysis based on Markov Chain Monte Carlo methods are presented in the Appendix.

\section{Mechanisms of Neutrinoless Double-Beta Decay}
\label{sec:mechanisms}

The standard mass mechanism for $0\nu\beta\beta$ decay assumes that the light active neutrinos $\nu_{i,L}$ are Majorana fermions, with a mass term $m_i \overline{\nu_{i,L}^C} \nu_{i,L}$ for each neutrino mass eigenstate $i = 1,2,3$. Here, $C$ denotes the action of the charge-conjugation operator. Evidently, the Majorana mass terms alter the lepton number by two units and owing to the left-handed nature of weak interaction, the observable quantity probed is the absolute value of the effective neutrino mass,
\begin{align}
\label{eq:eff-dbd-mass}
    \left| m_{\beta\beta} \right| =  \left| { \sum_i U^2_{ei} m_i } \right|,
\end{align}
where $U$ is the Pontecorvo–Maki–Nakagawa–Sakata matrix describing leptonic mixing. 

However, other Beyond-the-Standard-Model channels can induce a plethora of mechanisms for $0\nu\beta\beta$ decay if they incorporate lepton number violation (LNV), and, typically, generate light Majorana neutrino masses. At nuclear scales we can parameterize the exchange of heavy exotic states through effective 9-dimensional operators \cite{Pas2001},
\begin{align}
\label{srlagrangian}
    \mathcal{L}_\text{SR} = 
    \frac{G_F^2\cos^2\theta_C}{2m_p}\left(\epsilon_1 JJj 
                          + \epsilon_2 J^{\mu\nu}J_{\mu\nu}j 
                          + \epsilon_3 J^\mu J_\mu j 
                          + \epsilon_4 J^\mu J_{\mu\nu} j^\nu 
                          + \epsilon_5 J^\mu J j_\mu \right).
\end{align}
These are expressed as Lorentz-invariant products of two quark currents $J = \bar u(1 \pm \gamma_5)d$, $J^\mu = \bar u\gamma^\mu(1\pm \gamma_5)d$, $J^{\mu\nu} = \frac{i}{2}\bar u\left[\gamma^\mu, \gamma^\nu \right] (1 \pm \gamma_5) d$, and the $\Delta L = 2$ electron currents $j = \bar e(1\pm \gamma_5)e^C$, $j^\mu = \bar e\gamma^\mu(1\pm \gamma_5)e^C$, $j^{\mu\nu} = \frac{i}{2}\bar e\left[ \gamma^\mu, \gamma^\nu \right] (1 \pm \gamma_5) e^C$. The Fermi coupling constant $G_F$, the proton mass $m_p$ and the Cabibbo angle $\theta_C$ are factored out by convention to parameterize the interactions through dimensionless coupling constants $\epsilon_i$, which in addition to the type of Lorentz structure also include an implicit dependence on the chirality of the quarks and leptons involved. 

In general, all effective operators at dimension-7 and dimension-9 can contribute, in addition to the standard mass mechanism corresponding to the only possible dimension-5 Weinberg operator. The various contributions were calculated in \cite{Cirigliano:2017djv, Graf:2018ozy, Deppisch:2020ztt, Kotila:2021xgw}. We here concentrate on the interplay between the standard exchange of light Majorana neutrinos and a single short-range operator contribution, where the total $0\nu\beta\beta$ decay half-life $T_{1/2}$ in a given isotope may be expressed as 
\begin{align}
\label{eq:halflife-interference}
    T_{1/2}^{-1} = \frac{|m_{\beta\beta}|^2}{m_e^2} M_\nu^2 G_\nu 
                 + |\epsilon_I|^2 M_I^2 G_I 
                 + 2\,\text{Re}\left(\frac{m_{\beta\beta}}{m_e}\epsilon_I^*\right) 
                   M_\nu M_I G_{\nu I}.
\end{align}
Here, $G_\nu$, $G_I$, and $G_{\nu I}$ are the so-called phase space factors for the mass mechanism, short-range exotic contribution and interference, respectively. Despite their name, they represent the leptonic part of the transition matrix element and thus depend on the Lorentz structure and chirality of the lepton current. The NMEs $M_\nu$ and $M_I$ of the mass mechanism and short-range contribution, respectively, are by convention real numbers. These quantities can, however, have opposite signs, thus potentially affecting the third term describing interference. The electron mass $m_e$ is introduced to keep the dimensions of the NMEs and phase space factors uniform, i.e., the NMEs are dimensionless and the phase space factors are typically quoted in units of inverse years. 

Many ultraviolet-complete extensions of the SM induce the effective LNV operators in Eq.~\eqref{srlagrangian}. We choose a simplified version of the minimally-supersymmetric Standard Model with $R$-parity violation (RPV). In this model, a subset of the short-range operators is induced \cite{hirsch:1996ek},
\begin{align}
\label{srlagrangian-rpv}
    \mathcal{L}_\text{RPV} = \frac{G_F^2\cos^2\theta_C}{2m_p}\left(
          \epsilon^{RRL}_1 J_R J_R 
        + \epsilon^{RRL}_2 J_R^{\mu\nu} J_{R,\mu\nu}\right)j_L,
\end{align}
where we now explicitly denote the chirality of the currents, i.e., both the scalar and tensor quark currents are right-handed whereas the lepton current is left-handed.

The effective operators in Eq.~\eqref{srlagrangian-rpv} are induced by the exchange of heavy supersymmetric particles, namely squarks, gluinos, neutralinos and selectrons, under the presence of the RPV interactions of the form $\tilde e\bar u d$, $e\tilde{\bar u} d$ and $e\bar u \tilde d$ (i.e., with one superpartner denoted by the tilde). These operators alter lepton number by one unit, and thus $0\nu\beta\beta$ decay is induced at second order in the RPV interaction. The matching between the effective operators and the ultraviolet-complete model is given by \cite{hirsch:1996ek}
\begin{align}
\label{eq:rpv-matching}
    \epsilon_1^{RRL} = 
        \frac{8\pi}{9}
        \frac{\lambda^2_{111}\alpha_s}{\cos^2\theta_C}
        \frac{G_F^{-2}}{m_{\tilde q}^4}
        \frac{m_p}{m_{\tilde g}}, \qquad
    \epsilon_2^{RRL} = -\frac{1}{8}\epsilon_1^{RRL},
\end{align}
with the strong fine structure constant $\alpha_s = 0.127$ at $m_W$, the squark mass $m_{\tilde q}$ and the gluino mass $m_{\tilde g}$. The RPV interaction strength is denoted $\lambda_{111}$, with the indices indicating that the interaction couples to the first generation of (s)leptons and (s)quarks. The relations in Eq.~\eqref{eq:rpv-matching} assume degenerate squark masses and the so-called gluino dominance, i.e., diagrams mediated by gluinos dominate over those mediated by neutralinos. 

Because the effective operators in Eq.~\eqref{srlagrangian-rpv} have the same leptonic current as the standard mass mechanism, the phase space factors in Eq.~\eqref{eq:halflife-interference} are all equal, i.e., $G_\nu = G_I = G_{\nu I} = G^{(0)}_{11+}$ in the notation of \cite{Deppisch:2020ztt}. In such a scenario, the contributions from the mass mechanism and the short-range operators add coherently,
\begin{align}
\label{eq:halflife-interference-rpv}
    T_{1/2}^{-1} = G^{(0)}_{11+} 
        \left|
            \frac{m_{\beta\beta}}{m_e} M_\nu 
            + \epsilon_1^{RRL} M_1^{RRL}
            + \epsilon_2^{RRL} M_2^{RRL}
        \right|^2.
\end{align}
Using the relation between the short-range coupling constants in Eq.~\eqref{eq:rpv-matching}, we can simplify this to
\begin{align}
\label{eq:halflife-interference-rpv-simplified}
    T_{1/2}^{-1} = G^{(0)}_{11+} 
        \left|
            \frac{m_{\beta\beta}}{m_e} M_\nu 
            + \epsilon M_\text{SR}
        \right|^2,
\end{align}
with $\epsilon = \epsilon_1^{RRL}$ used from here on for simplicity, and the compound short-range NME $M_\text{SR} = M_1^{RRL} - M_2^{RRL}/8$.

The effective mass $m_{\beta\beta}$ and the short-range operator coupling $\epsilon$ are in principle complex numbers. For example, the phase of $m_{\beta\beta}$ arises from the Majorana phases in the leptonic mixing matrix $U$ through Eq.~\eqref{eq:eff-dbd-mass}. Separating magnitudes and phases, $m_{\beta\beta} \to \left| m_{\beta\beta} \right| e^{i\alpha_\nu}$, $\epsilon \to \left| \epsilon \right| e^{i\alpha_\text{SR}}$, we can write
\begin{align}
\label{eq:halflife-interference-rpv-phases}
    T_{1/2}^{-1} = G^{(0)}_{11+} 
        \left[
            \frac{ \left| m_{\beta\beta} \right|^2 } {m^2_e} M^2_\nu 
            + \left| \epsilon \right|^2 M^2_\text{SR}
            + 2 \left| m_{\beta\beta} \right| \left| \epsilon \right| M_\nu M_\text{SR} \cos(\alpha_\nu - \alpha_\text{SR})
        \right],
\end{align}
i.e., the half-life in general depends on three parameters; the magnitudes $\left| m_{\beta\beta} \right|$, $\left| \epsilon \right|$, and the relative phase $\alpha_\nu - \alpha_\text{SR}$. As already noted, the NMEs can be conventionally chosen to be real numbers. As discussed in Sec.~\ref{sec:nmes}, the NMEs $M_\nu$ and $M_{SR}$ have a relative minus sign. In the following, we make the simplifying assumption that the two contributions are relatively real. This amounts to choosing the relative phase to be $\alpha_\nu - \alpha_\text{SR} = 0$ or $\pi$. Equivalently, we can choose a positive real $m_{\beta\beta} \geq 0$ and real $\epsilon \lesseqgtr 0$. The half-life is then simply as in Eq.~\eqref{eq:halflife-interference-rpv-simplified}, but with complex-absolute dropped,
\begin{align}
\label{eq:halflife-analysismaster}
    T_{1/2}^{-1}(X) = G^{(0)}_{11+}(X) 
        \left[
            \frac{m_{\beta\beta}}{m_e} M_\nu(X) 
            + \epsilon M_\text{SR}(X)
        \right]^2.
\end{align}
Although we are omitting the relative phase between contributions, this captures the extreme cases where they constructively add up or destructively cancel.

\section{Interplay among multi-isotope half-life measurements}
\label{sec:concept}

In this paper we are interested in assessing the extent to which the multi-mechanism parameter space ($m_{\beta\beta}$ vs. $\epsilon$) can be probed by future \nubb-decay searches in different isotopes. This is indicated by the isotope label $X$ in Eq.~\eqref{eq:halflife-analysismaster} where we consider $X = $ \Ge, \Mo, \Xe. Both the NMEs and the phase space factors depend on the $0\nu\beta\beta$ decay isotope. In a given isotope, the parameters of interest, $m_{\beta\beta}$ and $\epsilon$, cannot be measured independently. For example, if $\epsilon = -(M_\nu / M_\text{SR})(m_{\beta\beta} / m_e)$, the contributions cancel exactly and $0\nu\beta\beta$ decay does not occur. While this amounts to an accidental cancellation as the particle physics parameters $m_{\beta\beta}$ and $\epsilon$ have to be fine-tuned against the nuclear physics parameters $M_\nu$ and $M_\text{SR}$, it must be resolved experimentally.

More generally, measuring a given half-life $T_{1/2}(X)$ in isotope $X$ corresponds to two parallel line solutions,
\begin{align}
    \epsilon = K(X) \cdot m_{\beta\beta} \pm C(X),
\end{align}
with the slope (positive in our scenario as $\text{sign}(M_\text{SR}) = - \text{sign}(M_\nu)$)
\begin{align}
    K(X) = - \frac{1}{m_e} \frac{M_\nu(X)}{M_\text{SR}(X)} > 0, \label{eq:kslope}
\end{align}
and the intercept
\begin{align}
    C(X) = \left[T_{1/2}(X) G^{(0)}_{11+}(X)\right]^{-1/2}.
\end{align}
We will take advantage of the fact that both the slope and intercept of these lines are isotope-dependent, which implies that the parameter space ($m_{\beta\beta}$ vs. $\epsilon$) can be constrained to a locus of two points using two $0\nu\beta\beta$ decay observations of distinct isotopes, and to a single point using three such observations.

\begin{figure}[t!]
    \centering
    \includegraphics[width=1\textwidth]{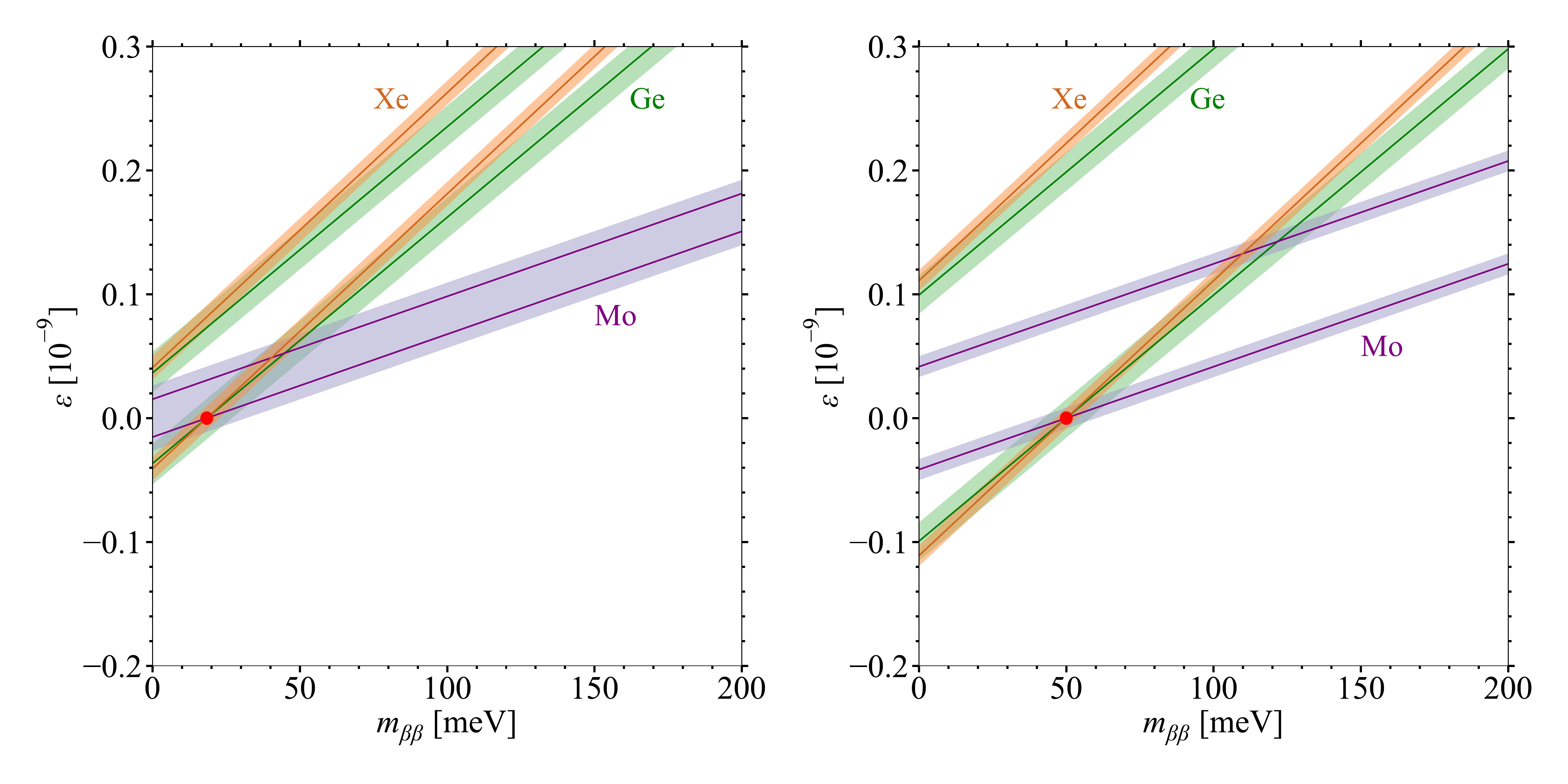}
    \caption{Parameter space allowed for the effective Majorana mass \mbb and short-range coupling $\epsilon$ assuming future measurements of the \nubb-decay half-life of \Xe (orange space), \Ge (green) and \Mo (purple). The measurements are assumed to have a central value corresponding to the half-life expected for $\epsilon = 0$ and  $m_{\beta\beta} = 18.4$~meV (left) or 50~meV (right), and to be affected by a $1\sigma$ statistical uncertainty as described in Sec.~\ref{statistics}. NME values are fixed to the best-fit estimates in Eq.~\eqref{eq:corrfit} and Table~\ref{tab:NMEvalues}. }
    \label{fig:individualmeasurements}
\end{figure}

This is illustrated in Fig.~\ref{fig:individualmeasurements}, showing the parallel lines for each isotope assuming the measurement of a half-life corresponding to the model parameters $m_{\beta\beta} = 18.4$ and 50~meV --- i.e., the minimal value allowed with the inverted mass ordering \cite{Agostini:2017jim} and a value just above the current limits, respectively --- and $\epsilon = 0$ --- i.e., no exotic contribution beyond the mass mechanism. The width of the bands corresponds to the expected $1\sigma$ statistical uncertainties of the half-life measurements, detailed in the next section. For $m_{\beta\beta} = 18.4$~meV, the half-lives and their uncertainties are $T_{1/2}(\Ge) = 1.2^{+2.9}_{-0.7} \times 10^{28}$~yr, $T_{1/2}(\Xe) = 4.0^{+4.7}_{-1.6} \times 10^{27}$~yr, $T_{1/2}(\Mo) = 1.7^{+5.5}_{-1.0} \times 10^{27}$~yr, and for $m_{\beta\beta} = 50$~meV they decrease to $T_{1/2}(\Ge) = 1.7^{+0.7}_{-0.4} \times 10^{27}$~yr, $T_{1/2}(\Xe) = 5.4^{+1.2}_{-0.9} \times 10^{26}$~yr, $T_{1/2}(\Mo) = 2.3^{+1.1}_{-0.7} \times 10^{26}$~yr. The NMEs values are considered perfectly known and correspond to the average of the estimates available in the literature. 

In a single given isotope, $m_{\beta\beta}$ and $\epsilon$ have a degeneracy as described and cannot be disentangled. However, when measurements are available in multiple-isotopes, the parameter space allowed is reduced to the overlap between bands. If measurements are available for two isotopes whose bands have significant different slopes (such as \Ge and \Mo, or \Xe and \Mo), the overlap creates two "islands". The first corresponds to a solution in which $\epsilon=0$, the second to a solution in which the contributions of nonzero $\epsilon$ and \mbb cancel. In the following, we will refer to the first as the "main solution", and to the second as the "satellite solution". 

While measuring $0\nu\beta\beta$ decay in \Ge and \Mo, or \Xe and \Mo provides two almost equiprobable solutions, measuring all three isotopes could in principle resolve the degeneracy. However, as the slopes of the \Ge and \Xe bands are similar, we expect the satellite solution to be only penalized rather than excluded.  We will quantify this effect through our Bayesian analysis in Sec.~\ref{sec:results}.

\section{Nuclear Matrix Elements and their Estimation}
\label{sec:nmes}

The NME for light-neutrino-exchange $0\nu\beta\beta$ decay may be defined by formally factoring lepton fields and Dirac algebra structures from weak vertices out of the decay amplitude, leaving a time-ordered product of left-handed charge-changing hadronic currents \cite{Engel:2016xgb},
\begin{align}
    M_\nu = \braket{f|J_L^{\mu}(x)J_L^{\nu}(y)|i}.
\end{align}
Inserting a complete set of intermediate nuclear states into the above expression allows one to factor the operator product into a product of single-operator matrix elements. However, even such simplified NMEs are impossible to evaluate exactly due to the many-body nature of the nuclear initial and final states $\ket{i}$, $\ket{f}$. Two approximations are therefore commonly made: the impulse approximation, where hadronic current operators are defined as collections of free-nucleon matrix elements; and the closure approximation, where the energies of intermediate states are taken to lie at a representative average energy. Together, these approximations are evidenced to introduce an uncertainty of $10\%$ or less~\cite{Engel:2016xgb}, and facilitate the decomposition of $M_\nu$ into Gamow-Teller~(GT), Fermi~(F) and tensor~(T) terms,
\begin{align}
    M_\nu = M_\nu^\text{GT} - \frac{g_V^2}{g_A^2}M_\nu^\text{F} + M_\nu^\text{T}.
\end{align}
Here, $g_V \sim 1$ and $g_A \sim 1.27$ are the effective vector and axial couplings in nuclear matter, and the terms are distinguished by their spin structures.

Given that external states $\ket{i},\ket{f}$ are not exactly known, many-body nuclear methods must be employed in connection with the above particle-exchange formalism. Details of competing many-body methods may be found in recent reviews \cite{Vergados2016, Engel:2016xgb, Dolinski:2019nrj}. The main classes of methods are: \emph{Energy-density Functional} (EDF) theory \cite{Rodriguez10, Vaquero14, Yao2015}; the \emph{Interacting Boson Model} (IBM) \cite{Arima1981,Iachello1987}; the \emph{Quasi-particle Random-Phase Approximation} (QRPA) \cite{Vogel1986,Engel1988,Hinohara2014}; and the \emph{Nuclear Shell Model} (NSM) \cite{Horoi16b,Menendez18,Coraggio20,Coraggio2022}.

Approximations in the many-body models and numerical methods are expected to introduce biases in the NME estimations. In particular, missing contributions to nuclear states and interactions are expected to consistently bias estimations on different isotopes made via the same many-body method, introducing a kind of ``systematic offset'': for example, the renormalization-enhanced contact operator recently identified \cite{Cirigliano:2021qko}; or the so-called quenching of the axial coupling $g_A$ \cite{Agostini:2022zub}. Such biases are expected to partially cancel when ratios between NME estimations of different isotopes are considered. Additionally, numerical and other approximations can inconsistently and randomly affect NME calculations, inducing a ``statistical uncertainty''. While a worldwide effort is ongoing to perform calculations in which these effects are minimized and theoretically bounded, based on \textit{ab initio} methods~\cite{Belley:2020ejd,Cirigliano2022} --- for the purposes of this work, we attempted to infer the magnitudes of these uncertainties from the (co)variances between NME values reported by different groups. We extract this information as a correlation fit on multiple isotopes, namely \Ge, \Mo and \Xe. 

Similar estimates of NME uncertainties were made earlier in QRPA by varying the axial coupling $g_A$, the short-range correlation and the configuration space \cite{Faessler2009}. Such an approach has been applied to probing exotic scenarios beyond the mass mechanism \cite{Fogli2009}.

\renewcommand{\arraystretch}{1.3}
\begin{table}[t!]
\centering
\begin{tabular}{llccc}
    \hline
    \multicolumn{2}{l}{Mass mechanism NMEs $M_\nu$} & \Ge & \Mo & \Xe \\
    \hline
    EDF & Vauqero et al. \cite{Vaquero14}                   & 5.55 & 6.59 & 4.77 \\
    EDF & Rodriguez, Martinez-Pinedo \cite{Rodriguez10}     & 4.60 & 5.08 & 4.20 \\
    REDF & Song et al. \cite{Song17}                        & 6.04 & 6.48 & 4.24 \\
    IBM-2 & Barea et al. \cite{Barea15}                     & 5.14 & 3.84 & 3.25 \\
    IBM-2 & Deppisch et al. \cite{Deppisch:2020ztt}         & 6.24 & 5.08 & 3.40 \\
    QRPA & Hyvarinen, Suhonen \cite{Hyvarinen15}            & 5.26 & 3.90 & 2.91 \\
    QRPA & Šimkovic et al. \cite{Simkovic18}                & 4.85 & 5.87 & 2.72 \\
    NSM & Corragio et al. \cite{Coraggio20,Coraggio2022}    & 2.66 & 2.24 & 2.39 \\ 
    \hline
    \multicolumn{2}{l}{Short-range NMEs $M_\text{SR}$} & \Ge & \Mo & \Xe \\
    \hline
    IBM-2 & Deppisch et al. \cite{Deppisch:2020ztt} ($M_1^{XX}$)    & $-5300$ & $-12400$ & $-3210$ \\
    IBM-2 & Deppisch et al. \cite{Deppisch:2020ztt} ($M_2^{XX}$)    & 174 & 189 & 96.1 \\
    \hline
    \multicolumn{2}{l}{Phase-space Factors $G^{(0)}_{11+}~[10^{-15}~\text{yr}^{-1}$]} & \Ge & \Mo & \Xe \\
    \hline
    & Deppisch et al. \cite{Deppisch:2020ztt} & 2.36 & 15.91 & 14.56 \\
    \hline
\end{tabular}
\caption{Light-neutrino exchange and short-range NMEs for $0\nu\beta\beta$ decay, calculated in different nuclear structure frameworks and adapted from \cite{Agostini2021}, and phase-space factors for the three isotopes under consideration.}
\label{tab:NMEvalues}
\end{table}

\begin{figure}[t!]
    \centering
    \includegraphics[width=\textwidth]{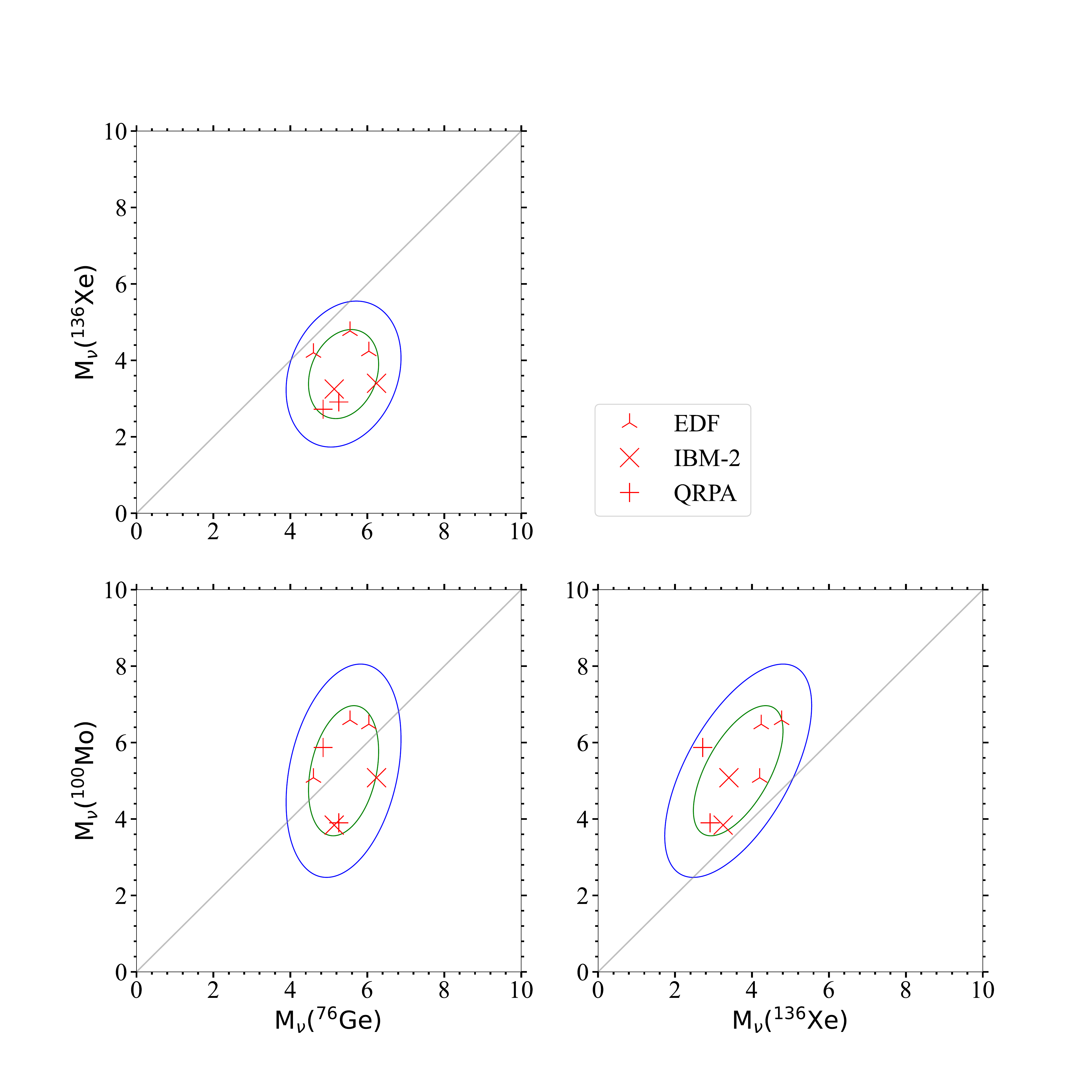}
    \caption{NMEs reported by EDF \cite{Rodriguez10, Vaquero14, Song17}, IBM-2 \cite{Barea15,Deppisch:2020ztt} and QRPA \cite{Hyvarinen15,Simkovic18}, with correlation ellipses corresponding to 1 and 2$\sigma$ confidence regions between isotope pairs of \Ge, \Mo, \Xe.}
\label{fig:NMEcorrelations}
\end{figure}

We build a probability distribution for the NME values across the three isotopes of interest for future searches by fitting the available estimations with a multivariate Gaussian distribution over the NMEs $\mathbf{m} = (M_\nu(\Ge), M_\nu(\Mo), M_\nu(\Xe))^T$,
\begin{align}
    \mathcal{N}(\mathbf{m}) = 
    \frac{1}{\sqrt{(2\pi)^3\,\text{det}(\Sigma)}}
    \exp\left[
        - \frac{1}{2}\left(\mathbf{m} - \boldsymbol{\mu}\right)^T \cdot \Sigma^{-1} \cdot \left(\mathbf{m} - \boldsymbol{\mu}\right)
    \right].
\label{gaussian}
\end{align}
The NME mean values $\boldsymbol{\mu} = \text{mean}(\mathbf{m}_i)$ and the covariance matrix $\Sigma = \text{cov}(\mathbf{m}_i)$ were obtained by fitting to NME calculations from the EDF, IBM-2, and QRPA frameworks provided in \cite{Agostini2021} and re-collated in Table~\ref{tab:NMEvalues}, where only calculations covering all three isotopes Ge, Mo, and Xe are included,
\begin{align}
    \boldsymbol{\mu} = \begin{pmatrix}
        5.383 \\ 5.263 \\ 3.641
        \end{pmatrix}, \qquad
    \Sigma = \begin{pmatrix} 
        0.361 & 0.200 & 0.102 \\ 
        0.200 & 1.260 & 0.527 \\ 
        0.102 & 0.527 & 0.590
    \end{pmatrix}.
    \label{eq:corrfit}
\end{align}
Two-dimensional projections of the data and the resulting $1$, $2$, and $3\sigma$ uncertainty ellipses are shown in Fig.~\ref{fig:NMEcorrelations}. The comparative imprecision in estimating $M_\nu(^{100}Mo)$ the Mo NME is apparent from these fits. The strength of correlation between Mo and Xe estimates also visibly outpaces the pairings involving Ge.

For the short-range NMEs, we make use of the IBM-2 calculations for $M_1^{XX}$ and $M_2^{XX}$ provided in Table~VI of \cite{Deppisch:2020ztt}, and re-collated in Table~\ref{tab:NMEvalues}. Since $|M_2^{XX}| \ll |M_1^{XX}|$, we only consider $M_1^{XX}$. As no other consistent set of calculation exists for short-range NMEs, we are prevented from applying the same fit procedure to identify correlated uncertainties. Instead, we assume that the relative uncertainty on short-range NMEs is the same as on our standard long-range NMEs. That is, for each isotope pair, we re-scale the fitted correlation matrix for $M_\nu$ to be centered at the short-range values (equivalently one could rescale all $\mathbf{m}_i$ by the same factor), 
\begin{align}
    \Sigma_\text{SR} = 
    \text{diag}\left(\frac{M_\text{SR}}{\mu}\right) 
    \cdot \Sigma \cdot 
    \text{diag}\left(\frac{M_\text{SR}}{\mu}\right) 
    = \begin{pmatrix} 
        0.350 & 0.464 & 0.089 \\ 
        0.464 & 6.994 & 1.095 \\ 
        0.089 & 1.095 & 0.459 
    \end{pmatrix} \cdot 10^6.
\end{align}
We then construct a multivariate Gaussian distribution $\mathcal{N}(\mathbf{m}_\text{SR})$ of the correlated short-range NMEs $\mathbf{m}_\text{SR} = (M_\text{SR}(\Ge), M_\text{SR}(\Mo), M_\text{SR}(\Xe))^T$. Correlation between the light-neutrino-exchange NMEs $\mathbf{m}_\nu$ and heavy NMEs $\mathbf{m}_\text{SR}$ is neglected, such that the joint NME probability distribution can be taken as the product $\mathcal{N}(\mathbf{m}_\nu) \cdot \mathcal{N}(\mathbf{m}_\text{SR})$

We lastly note that while this fit-based estimate of uncertainties and central values does not affect our exploratory study, this approach should be taken with a grain of salt. Proper uncertainty estimations should be integrated into many-body calculations, and any future analysis following our methodology should make use of those estimates together with improved constraints on contact-terms and  
"$g_A$-quenching" or similar effects. 

\section{Combination of Future Experimental Data} 
\label{statistics}

We model the outcome of future $0\nu\beta\beta$ decay searches using an effective counting analysis with a fixed background expectation~\cite{Agostini:2022zub}. Nowadays, all experiments rely on sophisticated multivariate analyses that are needed to constrain the background. However, the sensitivity is typically dominated by a rather small part of the multidimensional parameter space in which the signal-to-background rate is maximal. In this part of the parameter space, the number of signal and background events follows a Poisson distribution whose expectation can be expressed in terms of sensitive exposure $\cal{E}$ and sensitive background $\cal{B}$ for a given isotope~\cite{Agostini:2022zub},
\begin{align}
    \lambda_{0\nu} = 
        \ln 2 \cdot N_a \cdot {\cal E} \cdot m_\text{iso}^{-1} \cdot 
        T_{1/2}^{-1}(m_{\beta\beta}, \epsilon, M_\nu, M_\text{SR})
    \quad \text{and}\quad
    \lambda_{\text{bkg}} = \cal{B}\cdot\cal{E}.
\end{align}
Here, $N_a = 6.022 \cdot 10^{23}$ is Avogadro's number and $m_\text{iso}$ is the molar mass. The expected number of signal events depends on $m_{\beta\beta}$, $\epsilon$ and the NMEs through the half-life as defined in Eq.~\eqref{eq:halflife-analysismaster}. We take the sensitive background and exposure values after ten years of operations from \cite{Agostini:2022zub}. These values, listed in Table~\ref{tab:experiments}, combined with the simplified counting analysis, provide experimental sensitivities within 10\% of those quoted by the collaborations. 

We combine data from future experiments in a global likelihood given by the product of Poisson terms, one for each experiment with associated isotope,
\begin{align}
\label{eq:poisson}
    P(n_\text{obs}|X) = 
    \text{Poisson}(n_\text{obs}|\lambda_{0\nu} + \lambda_\text{bkg}) \simeq \frac{(\lambda_{0\nu} + \lambda_\text{bkg})^{n_\text{obs}} \cdot e^{-(\lambda_{0\nu} + \lambda_\text{bkg})}}{\Gamma(n_\text{obs}+1)},
\end{align}
and multivariate distributions $\cal{N}(\mathbf{m})$ and $\cal{N}(\mathbf{m}_\text{SR})$ describing the uncertainties and correlations of the NMEs,
\begin{align}
    {\cal L}(m_{\beta\beta}, \epsilon, \mathbf{m}, \mathbf{m}_\text{SR}) = 
        P(n_\text{obs}|\Ge) \cdot 
        P(n_\text{obs}|\Mo) \cdot 
        P(n_\text{obs}|\Xe) \cdot 
        {\cal N}(\mathbf{m}) \cdot 
        {\cal N}(\mathbf{m}_\text{SR}).
\end{align}
For the purpose of estimating sensitivities, we assume an Asimov data set, i.e., we set the number of observed events equal to the summed expectation for background and signal, $n_\text{obs} = \lambda_{0\nu} + \lambda_\text{bkg}$. In addition, to mitigate probability jumps due to the discrete nature of the Poisson distribution, we smooth it using its continuous interpolation with the gamma function in Eq.~\eqref{eq:poisson}~\cite{Agostini:2022zub}.

We consider the three experiments mentioned earlier, which are already at the conceptual design stage and can start construction as soon as funding become available: LEGEND-1000 (\Ge); CUPID (\Mo); and nEXO (\Xe). Their isotopic molar mass, sensitive exposure, and sensitive background parameters are listed in Table~\ref{tab:experiments}, alongside the expected number of events for signal and background, and discovery sensitivities as calculated in \cite{Agostini:2022zub}.
\begin{table}[t!]
\centering
\begin{tabular}{l|ccc}  
\hline
Experiment & LEGEND-1000 & CUPID & nEXO \\
\hline
Isotope    & \Ge         & \Mo   & \Xe  \\
$m_\text{iso}$ [g/mol]	&	 75.74   &  177.57   & 135.71  \\
$\mathcal{E}$ [$\text{kg}\cdot\text{yr}$]	&	 6632  &  1715   & 18618 \\  
$\mathcal{B}$ [$1/(\text{kg}\cdot\text{yr})$]	&	 $6.4 \times 10^{-5}$  &  $2.3 \times 10^{-3}$  & $2.9 \times 10^{-4}$ \\
\hline
$\lambda_{0\nu} (18.4~\text{meV})$ & \phantom{0}3.4 & \phantom{0}2.4 & \phantom{0}19.3 \\
$\lambda_{0\nu} (50.0~\text{meV})$ & 22.4 & 17.6 & 111.2 \\
$\lambda_{\text{bkg}}$ & \phantom{0}0.4 & \phantom{0}4.0 & \phantom{00}5.5 \\
$T^{3\sigma}_{1/2}$ [yr] & $1.3\times 10^{28}$ & $1.1\times 10^{27}$ & $7.5\times 10^{27}$ \\
$m^{3\sigma}_{\beta\beta}$ [meV] & $[9,21]$ & $[12,34]$ & $[6,27]$ \\
\hline
\end{tabular}
\caption{Parameters modelling the likelihood for LEGEND-1000, CUPID and nEXO as count-based experiments: molar mass $m_\text{iso}$, effective exposure $\mathcal{E}$ and background rate $\mathcal{B}$. For these setups, also listed are: the expected number of $0\nu\beta\beta$ signal events $\lambda_{0\nu}$ for $m_{\beta\beta} = 18.4$~meV and $50$~meV, the expected number of background events $\lambda_\text{bkg}$, the resulting $3\sigma$ discovery sensitivity $T_{1/2}^{3\sigma}$ and the corresponding sensitivity $m^{3\sigma}_{\beta\beta}$ on the effective mass, expressed as a range due to NME uncertainties. The values are taken from \cite{Agostini:2022zub}.}
\label{tab:experiments}
\end{table}

The parameters of interest for this analysis are $m_{\beta\beta}$ and $\epsilon$, and any statistical inference on them will require marginalization (profiling) over the NME values that act as nuisance parameters.  When the expected number of signal events is large and above the background level, and the uncertainties on the NME can be neglected, the likelihood becomes a product of independent normal distributions, and confidence intervals for $m_{\beta\beta}$ and $\epsilon$ can be extracted using a chi-square approximation. However, to properly fold NME uncertainties and correlations 
more sophisticated methods are called for.
This led us to adopt a full Bayesian analysis to derive posterior probability distributions on $m_{\beta\beta}$ and $\epsilon$, marginalizing over the NME values folded into the analysis as nuisance parameters with prior distribution given by Eq.~\eqref{eq:corrfit}. Flat priors are assumed for physics parameters $m_{\beta\beta}$ and $\epsilon$, although other choices are in principle possible; for $m_{\beta\beta}$, a comparative analysis of flat, logarithmic, and least-informative priors may be found in \cite{Deppisch2021}. The detailed theory and setup of our framework based on Markov Chain Monte Carlo methods is described in the appendix. 

The strong dependence of results on the size of the Markov chain proposal distribution was an obstacle to adequately sampling the parameter space. This dependence was explainable by noting the shape of the posterior distribution on $m_{\beta\beta}$ and $\epsilon$ is roughly that of a cone, narrowing sharply around the point which corresponds to the `true' values of the two parameters. Especially in high-dimensional parameter spaces, a narrow proposal distribution will excel at probing the sharp end of the cone, but converge much more slowly throughout the wide end; conversely, a wide proposal distribution will converge well throughout the wide end, but may rarely generate points in the sharp end or in any other small features of the distribution. For this reason and to standardize our results across scenarios, we followed the heuristic that $25\%$ is a desirable total acceptance rate, and in each case adjusted the width of our proposal distribution until this acceptance rate was achieved. The one exception occurred in multimodal scenarios where the primary and secondary peaks were so broadly disconnected that the above tuned proposal width could not thermalize across peaks on a practical timescale. Although multiple Markov chains would then precisely probe the posterior distribution shape for each disconnected peak, the likelihood ratio between peaks would be erroneously reported as unity. In such cases, we instead chose the minimal proposal width such that thermalization across peaks could be achieved within $10^7$ samples, which typically resulted in acceptance rates near $1\%$.

\section{Results}
\label{sec:results}

In this section, we apply our Bayesian analysis to explore the capability of future \nubb decay experiments to constrain \mbb and  $\epsilon$, assuming that half-life measurements are available for either two or three isotopes. We consider two extreme theory scenarios: \nubb decay mediated entirely by the light-neutrino-exchange ($\epsilon = 0$, see Sec.~\ref{scenario1}) or only by more exotic channels ($\mbb= 0$, see Sec.~\ref{scenario2}). These extreme scenarios are chosen to illustrate the interplay of the different NME values in comparatively stronger long- and short-range mechanism, and can be interpolated to obtain results for any intermediate scenario with mixed contributions. Finally, in addition to our baseline probability distribution for the NME values discussed in Sec.~\ref{sec:nmes}, we repeat our calculation assuming a 10-times narrower distribution (i.e., reducing the variance of the distributions by a factor of 10) to show how the situation would change when the uncertainties on NME calculations become sub-dominant compared to the those on the decay half-life measurements.

\subsection{Standard Mass Mechanism}\label{scenario1}

\begin{figure}[t!]
    \centering
    \includegraphics[width=1.0\textwidth]{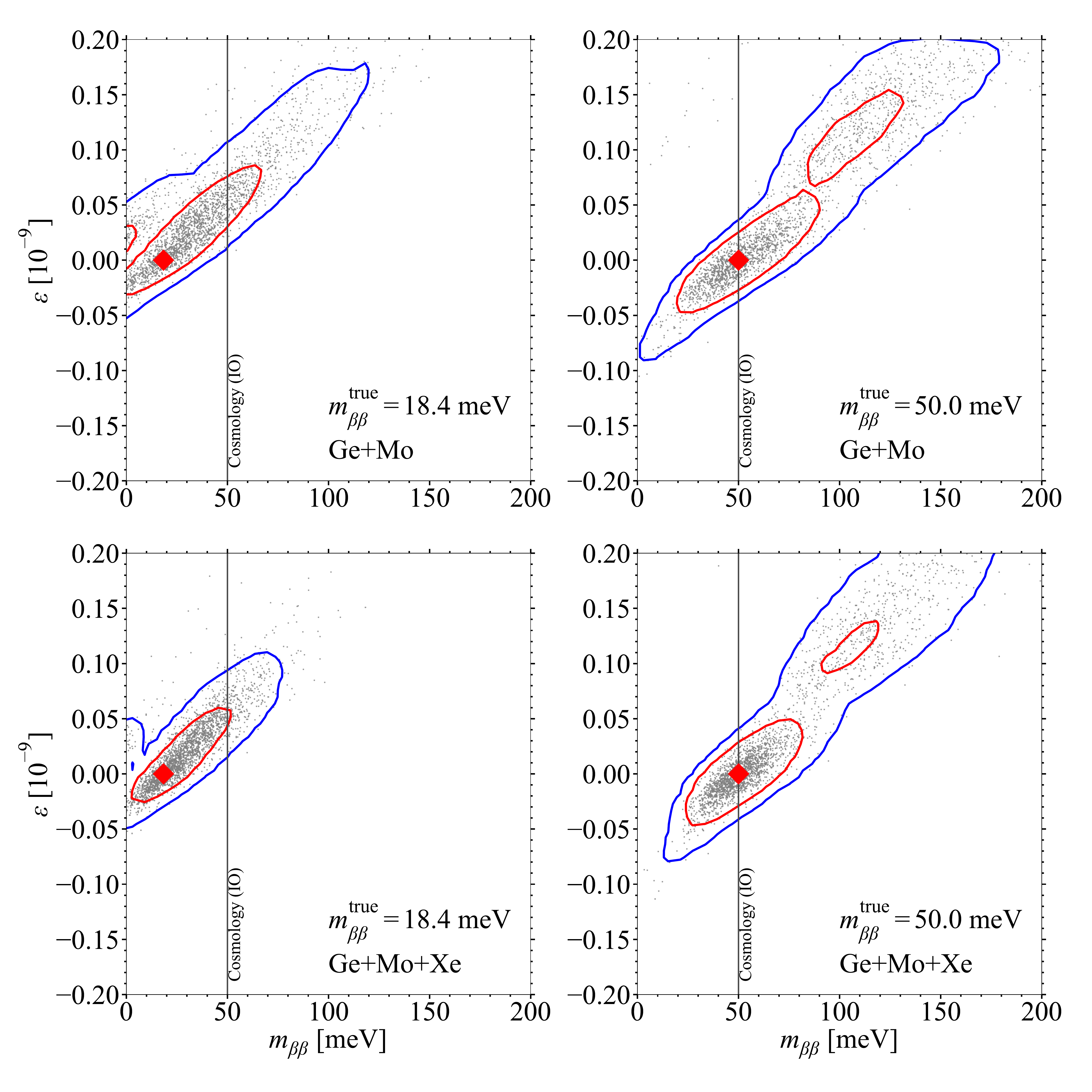}
    \caption{Bayesian posterior probability distributions and credible regions, with current NME uncertainties, for half-life values corresponding to $m_{\beta\beta} = 18.4$~meV (left column) and $m_{\beta\beta} = 50$~meV (right column), and isotopes Ge+Mo (top row) and Ge+Mo+Xe (bottom row). $68\%$ and $95\%$ probability credible regions are plotted in red and blue, respectively. The present cosmology bound on $m_{\beta\beta}$ from the inverted neutrino mass ordering (IO), described in the text, is plotted as a vertical line.}
    \label{fig:mcmclikes_m}
\end{figure}

Figure~\ref{fig:mcmclikes_m} shows how future measurements on the \nubb decay half-life of two (Ge+Mo, top row) and three isotopes (Ge+Mo+Xe, bottom row) can constrain the value of \mbb and $\epsilon$. The allowed parameter space is presented as $68\%$ and $95\%$-probability smallest credible regions marginalized over the NME nuisance parameters, obtained through our Bayesian analysis using the experimental sensitivities and NME correlations discussed in the previous sections. The figure shows the allowed parameter space assuming that experiments measure the half-life expected for 
$\epsilon = 0$ and \mbb equal to either 18.4~meV --- i.e., the lowest value for inverted-ordered neutrinos~\cite{Agostini:2017jim} (see left column) --- or 50~meV --- i.e., the largest value considering current cosmology bounds~\cite{Deppisch:2020ztt} (see right column).
Compared to the simplified picture illustrated in Fig.~\ref{fig:individualmeasurements}, our Bayesian analysis treats the NME values as correlated nuisance parameters affected by uncertainties. As a consequence, the parameter space allowed for a single isotope is no longer a straight pair of bands, but develops a more complicated shape which spreads outwards for large $m_{\beta\beta}$ values. This cone-like spreading is generated by the uncertainty on the slope $K$ defined in Eq.~\eqref{eq:kslope}, which is no longer fixed. 
When data from multiple isotopes are combined, the resultant allowed parameter space is now given not only by the overlap of the spaces for the single-isotopes, but further reflects a variable probability density which decreases for large values of \mbb.  

This has a strong impact on the "main" and "satellite" solutions discussed in the Sec.~\ref{sec:concept}, appearing when the half-life is large and the two solutions can be resolved (see the right panels corresponding to $\mbb=50$~meV). In this case, the two solutions are no longer equiprobable, and the satellite solution is penalized.

For both half-life assumptions, the improvement due to observation in a third isotope (Xe) manifests as an overall narrowing of the credible intervals and a further suppression of parameter space at large $\epsilon$ and \mbb values, decreasing the tail of the distribution and suppressing the satellite solution. The key property of these restrictions on the parameter space is their dependence on the slope of the band of equivalent half-life for the third isotope. \Xe has a similar slope to that of \Ge for our exotic contribution model, but quite distinct from that of \Mo; three isotopes of similar slope would generate a much broader allowed region of parameter space, while three isotopes all with distinct slopes could more effectively suppress the satellite solution and produce a narrower credible interval. We note that including a third isotope in our analysis is particularly critical for the exclusion of vanishing \mbb values.

\begin{figure}[t!]
    \centering
    \includegraphics[width=1.0\textwidth]{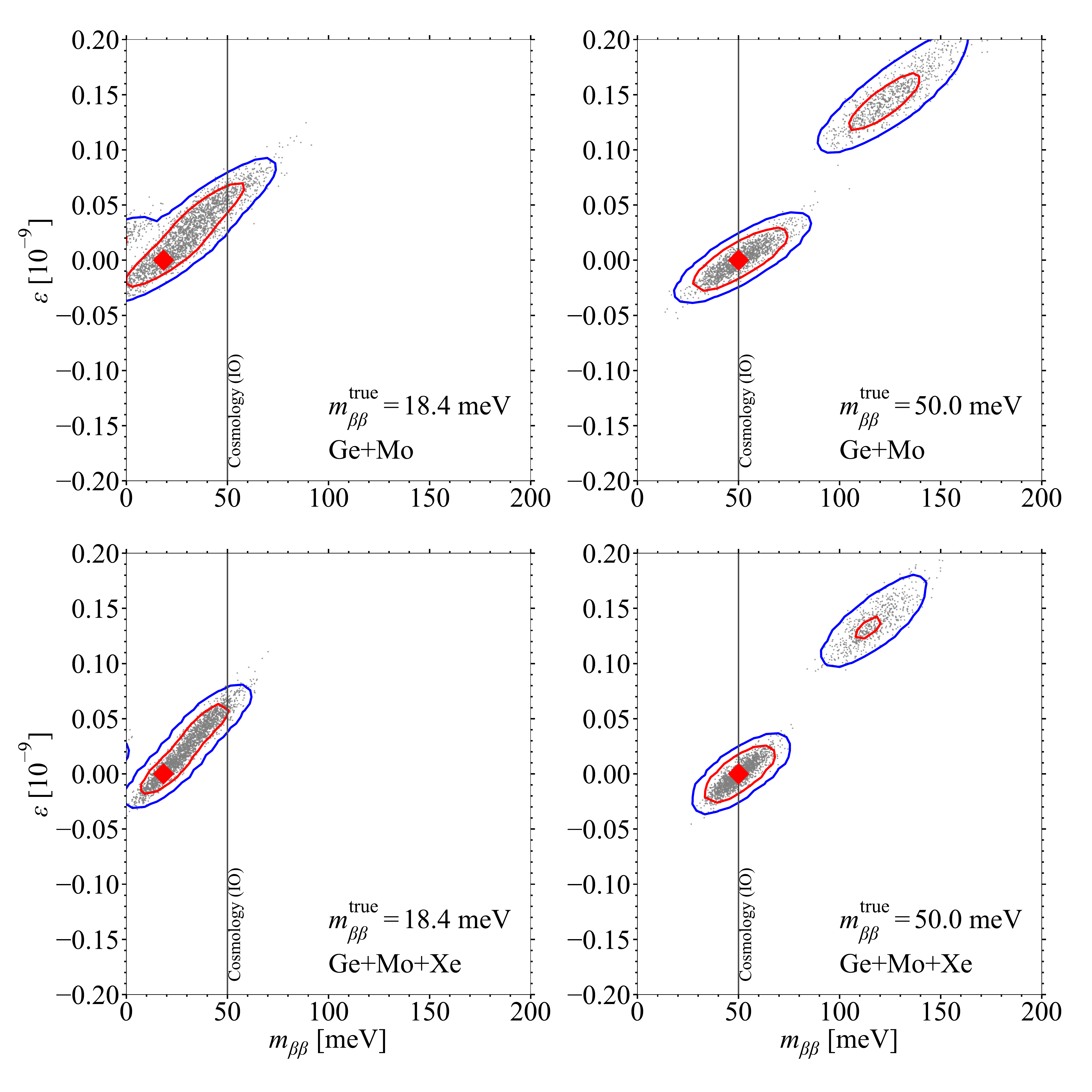}
    \caption{As Fig.~\ref{fig:mcmclikes_m}, but with reduced NME uncertainties.}
    \label{fig:mcmclikes_m_fut}
\end{figure}

If we repeat the analysis assuming a 10-times narrower NME probability distribution,
the credible regions contract for all considered scenarios (see Fig.~\ref{fig:mcmclikes_m_fut}).
Notably, the $95\%$-probability smallest credible regions no longer conjoin for half-life values corresponding to $m=50$~meV, nor does the tail of the main solution any longer cross the $m_{\beta\beta}=0$-axis. Consequently, the power of an observation to measure $\epsilon = 0$ and to exclude $m_{\beta\beta}=0$ is strengthened. This prospect strongly motivates further efforts within the nuclear theory community to improve the accuracy of the NME estimations.

\begin{figure}[t!]
    \centering
    \includegraphics[width=1.0\textwidth]{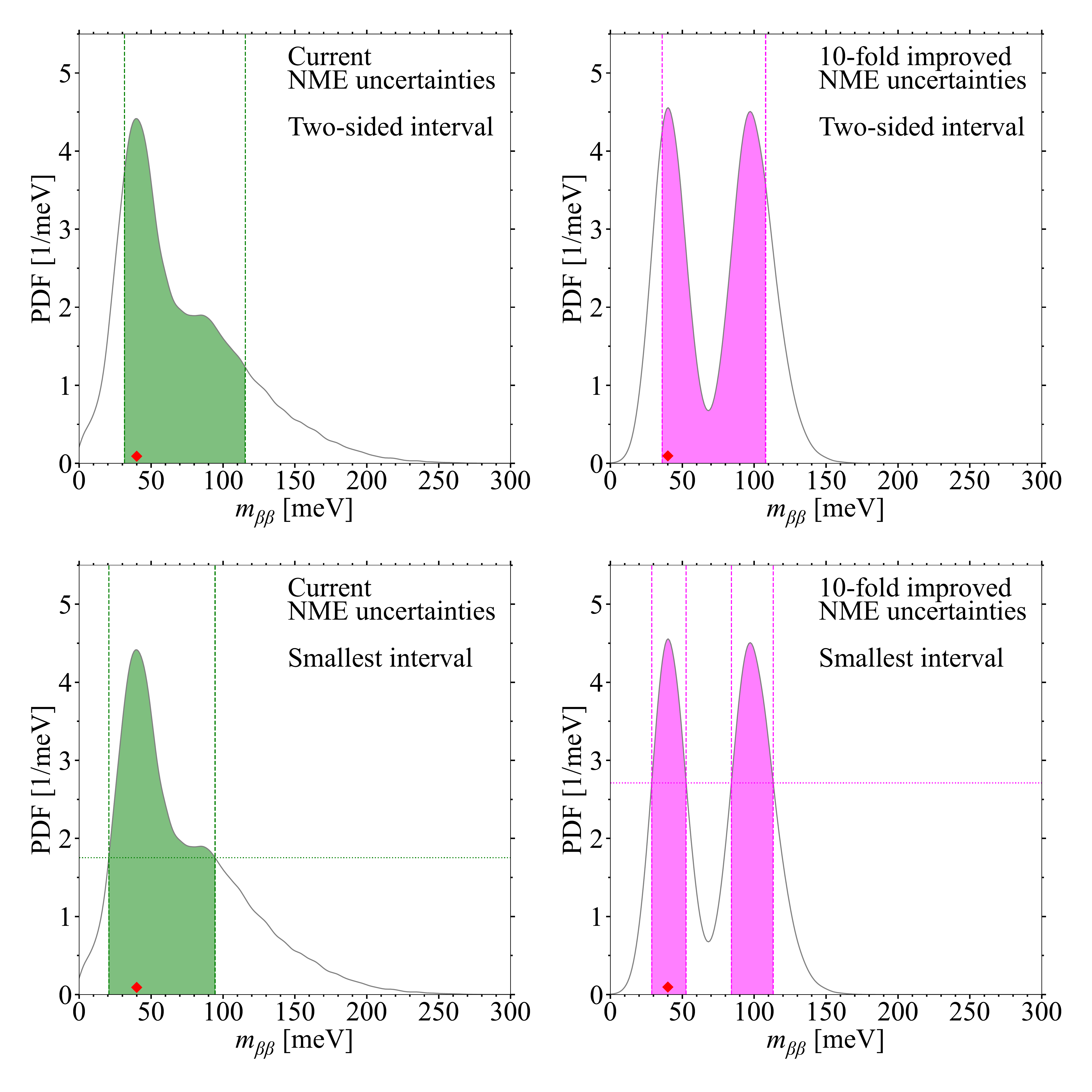}
    \caption{Comparison between central (top) and smallest (bottom) credible intervals, for both current and reduced NME uncertainties for half-life values corresponding to $m_{\beta\beta}=40$~meV, $\epsilon=0$. Note that the smallest intervals correspond to those shown in the left panel of Fig.~\ref{fig:heatmap2_m}.}
    \label{fig:interval_comparison}
\end{figure}

\begin{figure}[t!]
    \centering
    \includegraphics[width=.49\textwidth]{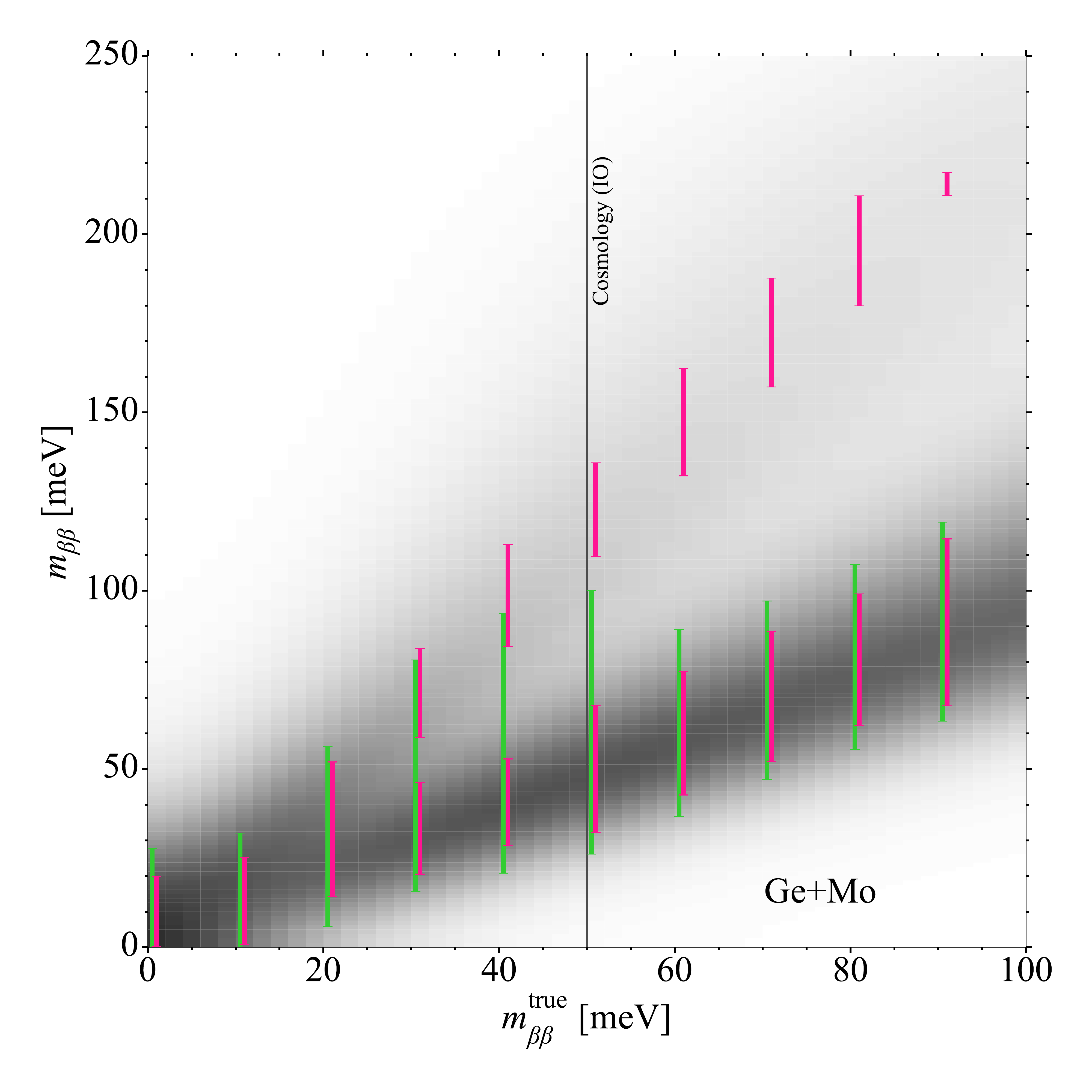}
    \includegraphics[width=0.49\textwidth]{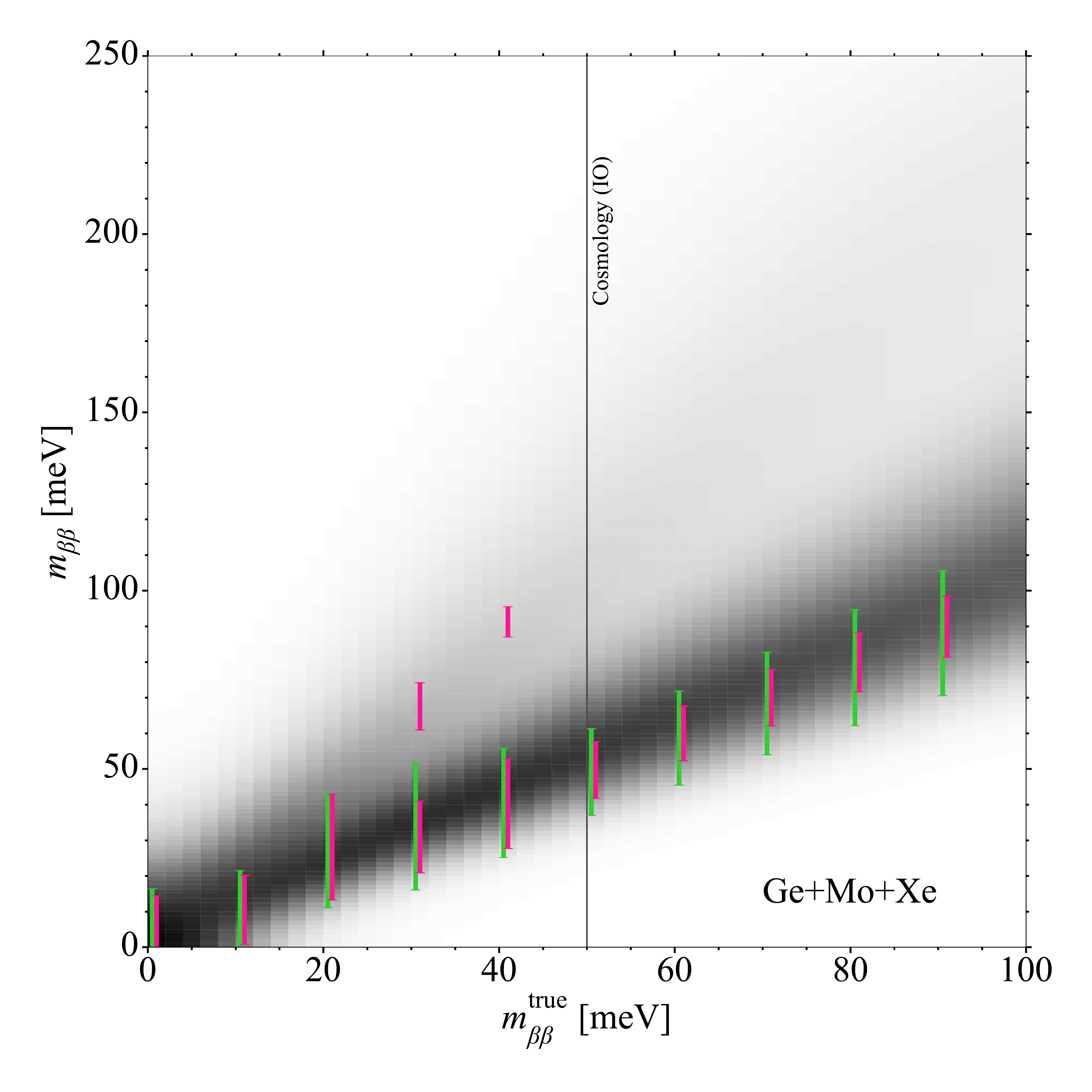}
    \caption{$68\%$ credible intervals over $m_{\beta\beta}$ for different $m^{\text{true}}_{\beta\beta}$, computed as in the bottom row of Figure \ref{fig:interval_comparison}, for our baseline NME probability distribution (green) and a 10-times narrower one (magenta) as discussed in the text, in the case of Ge+Mo observation (top) and Ge+Mo+Xe observation (bottom). To illustrate the primary and secondary bands, marginalized posterior distributions are plotted as vertical histograms in gray.}
    \label{fig:heatmap2_m}
\end{figure}

\begin{figure}[t!]
    \centering
    \includegraphics[width=.49\textwidth]{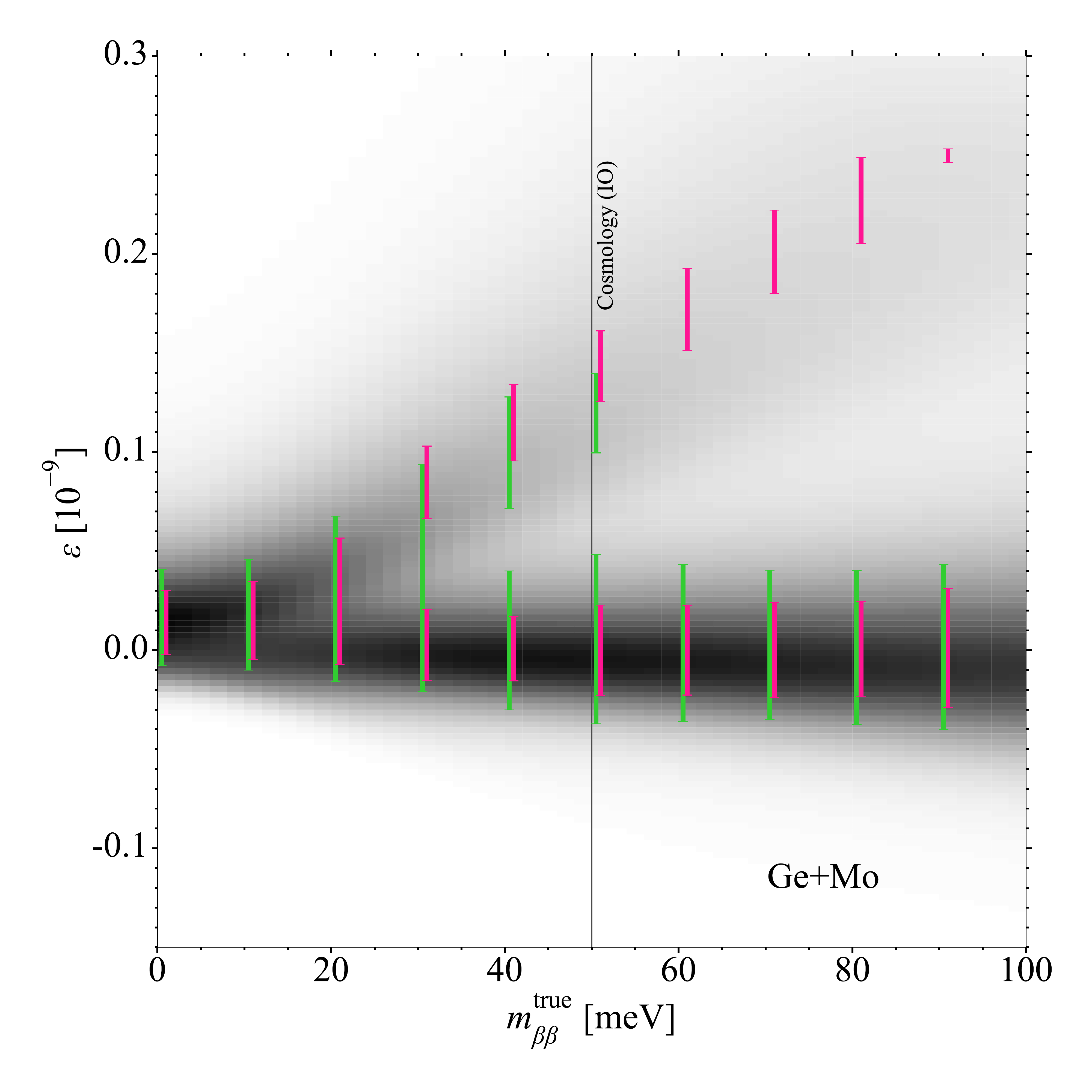}
    \hspace{2pt}
    \includegraphics[width=0.49\textwidth]{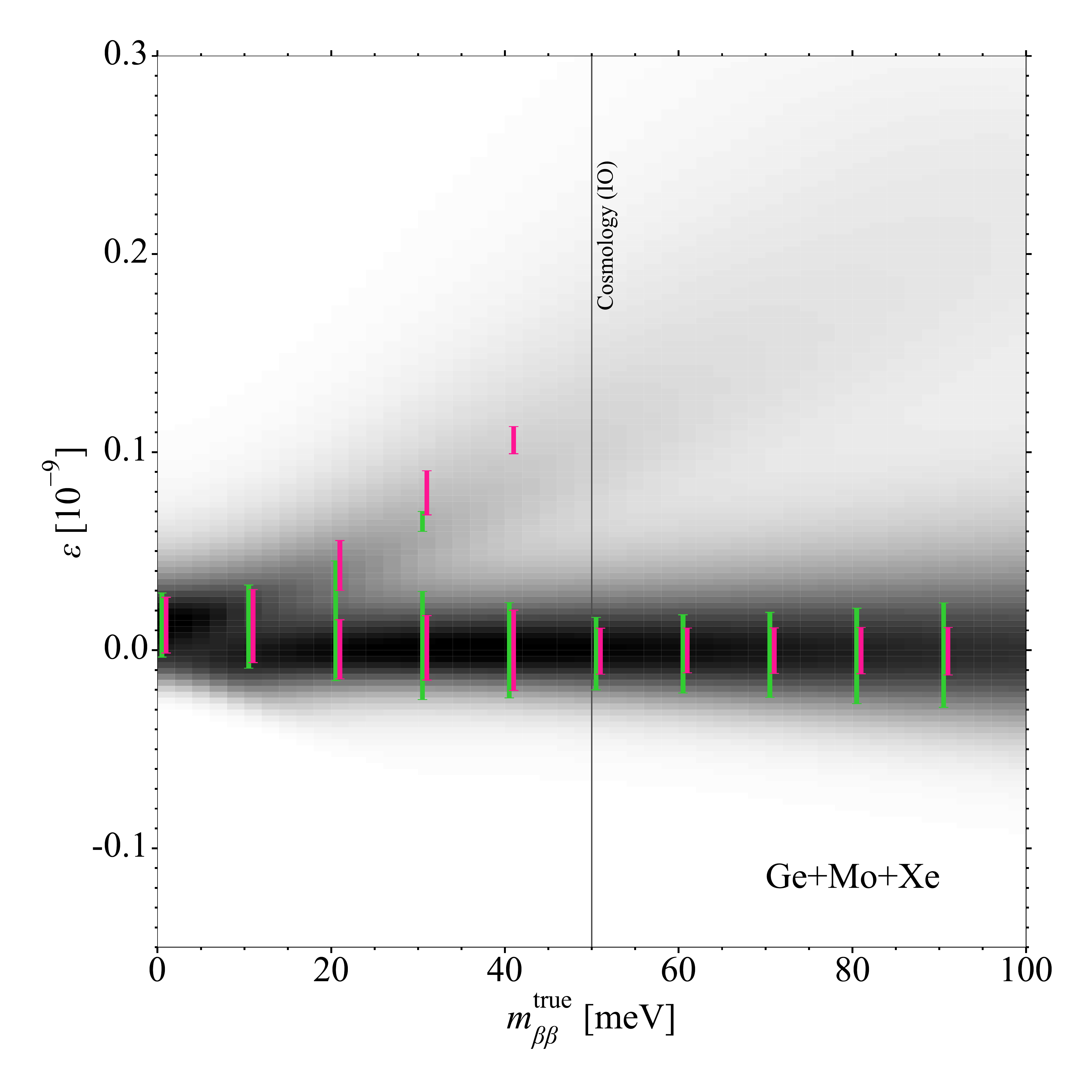}
    \caption{As Fig.~\ref{fig:heatmap2_m}, but marginalized over $\epsilon$.}
    \label{fig:heatmap2_e}
\end{figure}

At this point we mention a subtlety of constructing credible intervals in the case of a multimodal marginalized posterior distribution. Credible intervals constructed by rejecting the left- and right-most probability areas are shown in the top row of Fig.~\ref{fig:interval_comparison}. Most apparent in the case of reduced  NME uncertainties, but still important for our baseline NME uncertainties, is that this analysis fails to identify the significantly-rejected region around $50$~meV between the main and satellite solutions. 
Instead, smallest credible intervals constructed by finding the smallest --- possibly disjointed --- interval containing a given probability (see bottom row of Fig.~\ref{fig:interval_comparison}) clearly captures both the region of exclusion between the solutions and the relative suppression of the satellite solution. All credible intervals shown in Fig.~\ref{fig:heatmap2_m} and \ref{fig:heatmap2_e} are smallest intervals, which is furthermore analogous to the method by which all credible regions shown in Fig.~\ref{fig:mcmclikes_m} and \ref{fig:mcmclikes_e_fut} are computed.

The exemplar scenarios visualized in Fig.~\ref{fig:mcmclikes_m} and \ref{fig:mcmclikes_m_fut} correspond to the two potential half-life values expected for $\epsilon=0$ and $\mbb=18.4$ or $50$~meV. To show how these results should be interpreted for other values, we have taken the situation of pure light-neutrino-exchange $\epsilon=0$ and conducted a scan of half-life over $0 \leq m_{\beta\beta} \leq 100$~meV with intervals of $2$~meV. For each scanned value, we plot in Fig.~\ref{fig:heatmap2_m} the $68\%$-probability smallest credible interval (magenta bars). The smallest intervals are disconnected when the main and satellite solutions can be resolved, and the two solutions appear here as a primary band along the ``diagonal'' $m_{\beta\beta}=m^{\text{true}}_{\beta\beta}$ and a steeper secondary band. As $m^{\text{true}}_{\beta\beta}$ is increased, the secondary band tapers off as a direct consequence of the cone-spreading effect from NME uncertainties. However, this secondary band notably broadens the bounds placed on $m_{\beta\beta}$ by two-isotope observation for $m^{\text{true}}_{\beta\beta} \leq 50$~meV, and by three-isotope observation for $m^{\text{true}}_{\beta\beta} \leq 20$~meV. It is perhaps most clear in this visual how significantly the signal from a distinct third isotope improves the measurement situation, with NME uncertainties facilitating an overall contraction of the bounds placed on $m_{\beta\beta}$ by a factor of $\sim1.5$ regardless of $m^{\text{true}}_{\beta\beta}$. The figure also shows the $68\%$-probability central intervals build around the median of the \mbb marginalized posterior distribution. In this case the interval is always fully connected and allow to visually extract from the plot the value of upper or lower limits on \mbb.

While there is still a notable improvement in constraints from two- to three-isotopes when a narrower NME probability distribution is considered, the improvement is not so strong throughout the parameter space as with current NMEs. As for the individual example half-life values, this occurs because the narrower NME distribution weaken the broadening of the satellite solution, a broadening which the third isotope advantageously enhances; that is, the degeneracy between solutions requires more data to reject when the NMEs are more precisely known. The consequence is that a statistical rejection of the satellite solution may occur with our baseline NMEs  which is more challenging when narrowing their distribution, in particular between $30 \leq m^{\text{true}}_{\beta\beta} \leq 50$~meV as seen in Fig.~\ref{fig:heatmap2_m}.

The above procedure is now followed again but with marginalization over $\epsilon$, with results shown in Fig.~\ref{fig:heatmap2_e}. The primary band now runs along the horizontal $\epsilon = 0$, as $\epsilon^{\text{true}}=0$ has been enforced, while the secondary band is directed upwards (note that only samples $m_{\beta\beta} > 0$ are included in the marginalization). The secondary band again broadens the bounds placed on $\epsilon$ by 2-isotope observation for $m^{\text{true}}_{\beta\beta} \leq 50$~meV, and by 3-isotope observation for $m^{\text{true}}_{\beta\beta} \leq 20$~meV, and the signal from a distinct third isotope again improves the measurement situation throughout the scan.

We observe that if $\epsilon^{\text{true}}=0$ and $20$~meV~$< m^{\text{true}}_{\beta\beta} < 50$~meV, a third isotope provides the strongest possible benefit, in a factor of $2-3$ improvement in simultaneously constraining a measured $m_{\beta\beta}$ and excluding new physics in the form of a nonzero $\epsilon$. This region of course depends on the assumed uncertainties on both the light and heavy NMEs but points generally to the potential for a simultaneous three-isotope observation to decisively improve on two-isotope observations in the search for exotic mechanisms for $0\nu\beta\beta$~decay. The tendency of a narrower NME probability distribution to strengthen the degeneracy between solutions (while of course simultaneously improving overall constraining power) remains present, but is less notable in the construction of marginalized constraints over $\epsilon$ due to the reflection symmetry of the parameter space and consequent presence of two primary solutions at $\epsilon=0$. 

\subsection{Exotic Mechanism}\label{scenario2}
\begin{figure}[t!]
    \centering
    \includegraphics[width=1.0\textwidth]{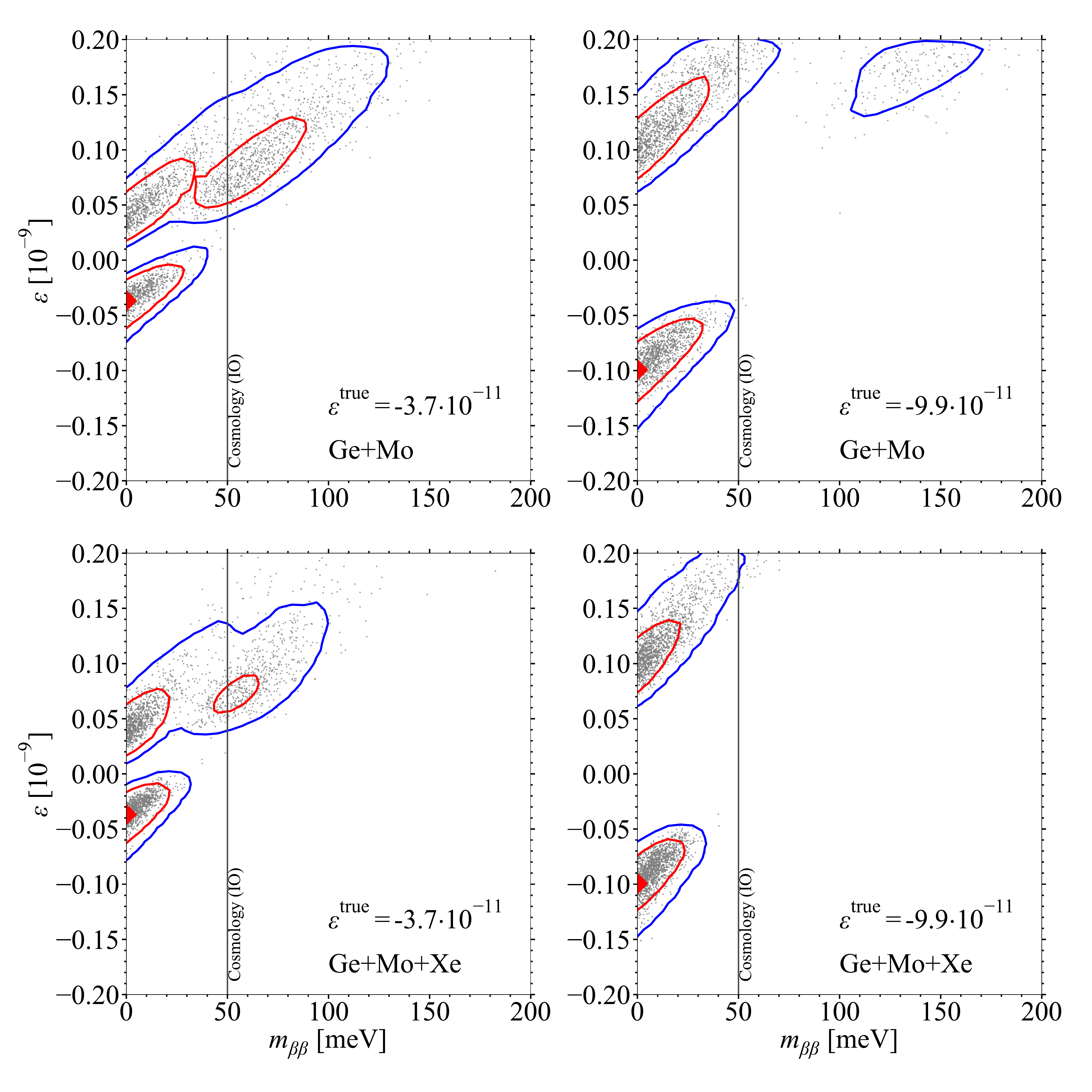}
    \caption{Bayesian posterior probability distributions and credible regions, with current NME uncertainties, for half-life values corresponding to $\epsilon = -3.68 \cdot 10^{-11}$ (left column) and $\epsilon = -9.94 \cdot 10^{-11}$ (right column), for isotopes Ge+Mo (top row) and Ge+Mo+Xe (bottom row). $68\%$ and $95\%$ probability credible regions are plotted in red and blue, respectively. The present cosmology bound on $m_{\beta\beta}$ from the inverted neutrino mass ordering (IO), described in the text, is plotted as a vertical line.}
    \label{fig:mcmclikes_e}
\end{figure}

We additionally consider an exotic scenario, where $0\nu\beta\beta$ decay is dominated by a short-range process parameterized by $\epsilon$. Figure~\ref{fig:mcmclikes_e} shows the credible regions constructed similarly to the those discussed in the previous section, but now for half-live values corresponding to $m_{\beta\beta} = 0$~meV and $\epsilon= -3.68 \cdot 10^{-11}$ in the left column, and with $\epsilon = -9.94 \cdot 10^{-11}$ in the right column. These values are selected as they correspond to the very same \Ge half-live used in the previous section. 

Although the half-life for \Ge has been held constant in considering these exotic physics scenarios, the half-life for \Mo has decreased from the pure light-neutrino exchange scenarios due to their smaller slope (and \Xe increased slightly). The result seen in Fig.~\ref{fig:mcmclikes_e} is an improved separation between the tow solutions, and an increased suppression of the satellite one, which lies farther along the single-isotope cone. We note that the two high-probability-density regions adjoining the $m_{\beta\beta}$-axis should be viewed as a single primary cluster bisected by that axis, by the reflection symmetry of the half-life in this parameter space. The most significant practical feature of this scenario is that $\epsilon = 0$ may be excluded, not just at the origin point, but after marginalization over $m_{\beta\beta}$ as well. Both exclusions are slightly enhanced by an observation in a third isotope (bottom row); the impact would be greater if there were a greater difference between the slopes of Ge and Xe. We also note that the positivity of all three isotope slopes leads to an asymmetric exclusion range for $\epsilon$, which is stronger in the positive direction than the negative direction, and that this holds regardless of the sign of $\epsilon$.

\begin{figure}[t!]
    \centering
    \includegraphics[width=1.0\textwidth]{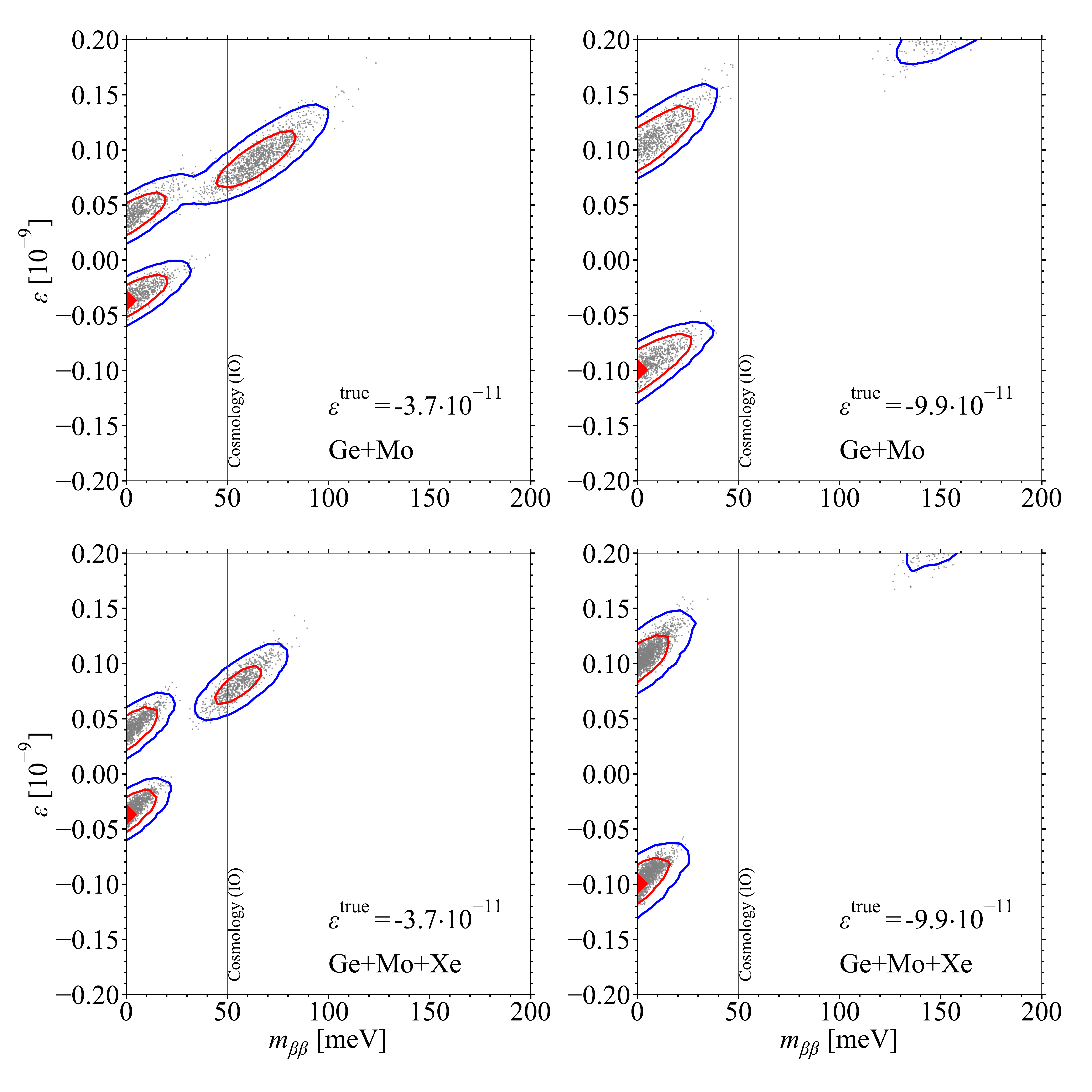}
    \caption{As Fig.~\ref{fig:mcmclikes_e} but with \textit{ten-times narrower} NME probability distributions.}
    \label{fig:mcmclikes_e_fut}
\end{figure}

Figure~\ref{fig:mcmclikes_e_fut} correspondingly shows the credible intervals obtained assuming a ten-times narrower NME probability distribution. As for the standard mass mechanism, a significant contraction of the credible regions is observed in line with reduced  NME uncertainties. However, in contrast to the standard mass mechanism, even for the smaller $\epsilon^{\text{true}} = -3.68 \cdot 10^{-11}$ the main and satellite solutions are disconnected. This is a significant realisation, for it implies a range of realistic scenarios, within current cosmology limits, where two or three observations of $0\nu\beta\beta$ decay are made and yet where it remains impossible to distinguish between a pure exotic mechanism (with unknown sign) and a mixed exotic-standard mechanism (with $\epsilon$ and $m_{\beta\beta}$ of the same sign).

\section{Conclusion}
\label{conclusion}

Observation of $0\nu\beta\beta$ decay is well-understood to probe the Majorana nature of light neutrinos. In addition, it is sensitive to other sources of lepton number violation, typically arising in New Physics scenarios above the electroweak scale that generate light Majorana neutrinos. If $0\nu\beta\beta$ decay is observed, it will be critical to distinguish between scenarios, most importantly between the standard mass mechanism of light-neutrino-exchange and a (more) exotic contribution. Specifically, we sought to describe how effectively $m_{\beta\beta}$ can be measured and the admixture of an exotic contribution can be constrained. 

Most exotic contributions, such as that induced in $R$-parity violating supersymmetry considered here, cannot be distinguished from the mass mechanism by measuring the total number of $0\nu\beta\beta$ decay events in a single isotope. Instead, drawing motivation from recent evidence that nuclear-structure correlations can break an otherwise-strong isotopic degeneracy in the ratio of mass-mechanism to exotic NMEs, we have demonstrated that observations in two isotopes with sufficiently different NME ratios (such as \Ge and \Mo, or \Xe and \Mo) can simultaneously constrain not only $m_{\beta\beta}$ but also the exotic mechanism, parameterised by an effective coupling constant $\epsilon$. 

An anomalous, satellite solution in $(m_{\beta\beta}, \epsilon)$ appears in this case, but it can be suppressed by an observation in a third isotope. We perform a numerical Bayesian analysis of planned $0\nu\beta\beta$ decay searches in isotopes \Ge, \Mo, and \Xe, namely LEGEND-1000, CUPID and nEXO, respectively. We use an averaged set of NMEs based on recent calculations in three different nuclear-structure models, including correlation of NMEs in different isotopes. We demonstrate the extent to which a third isotope improves on the joint constraints from two-isotope observations. We also demonstrate how a reduction of NME uncertainties will facilitate improved joint constraints in comparison to current NMEs. Both results are examined for minimal and maximal scenarios permitted by cosmology in the inverted neutrino mass ordering.

By examining the dependence of our constraints on $m_{\beta\beta}$ in the case of a pure light-neutrino-exchange mechanism, we demonstrate the practical effects of the satellite solutions for marginalized constraints on both $m_{\beta\beta}$ and $\epsilon$. For observation in  \Ge-\Mo or \Xe-\Mo, a notable transition between $68\%$-probability smallest credible intervals including both solutions to those including only the main one occurs around $m_{\beta\beta} = 50$~meV. However, with observation in \Ge, \Mo, and \Xe, this transition is pushed down to $30$~meV. Crucially, the impact of the NME uncertainties is not homogeneous throughout the parameter space; rather, they result in a cone-shaped spreading of the degenerate single-isotope bands as they extend further from the true parameter values. Therefore the satellite solution is, counter-intuitively, more readily suppressed by a third isotope with our baseline NME uncertainties in comparison to reduced ones. That said, observation in a third isotope appears to dominate the statistics, and even with reduced NME uncertainties the aforementioned transition occurs around $40$~meV. Beyond the satellite solution, reducing NME uncertainties by a factor of ten has the expected effect of narrowing the credible intervals marginalised over both $m_{\beta\beta}$ and $\epsilon$, typically by $20-30\%$.

Our results demonstrate the value of searching for $0\nu\beta\beta$ decay in multiple isotopes as a unique tool to distinguish decay mechanisms, and highlight the importance of a thorough treatment of the NME estimates, for which a proper uncertainty estimation will be crucial to interpret future discoveries.

\acknowledgments{The authors acknowledge support from the Science and Technology Facilities Council, part of U.K. Research and Innovation, Grant No. ST/P00072X/1, ST/T000880/1, ST/T004169/1 and ST/W00058X/1. G.~V.~G. also acknowledges support from the UCL Centre for Doctoral Training in Data Intensive Science funded by STFC, and from the UCL Overseas Research Scholarship / Graduate Research Scholarship. M.~A. acknowledges support from the UCL Cosmoparticle Initiative. The work of F.~F.~D. was performed in part at the Aspen Center for Physics, which is supported by National Science Foundation grant PHY-1607611. This work was partially supported by a grant from the Simons Foundation.}

\appendix

\section{Appendix: Markov Chain Monte Carlo Simulation}
\label{mcmc}

The Bayesian approach to statistics treats models as having associated degrees of belief, which for a model parameterised by continuous variables $\boldsymbol{\theta}$ manifests as a normalized probability distribution over $\boldsymbol{\theta}$. The goal of a practitioner is to make use of observed data $\boldmath{X}$ to refine a prior belief distribution $\pi(\boldsymbol{\theta})$ into a posterior belief distribution $p(\boldsymbol{\theta})$. Defining a likelihood distribution over the data space $L_{X}(\boldsymbol{\theta}) \equiv P(X|\boldsymbol{\theta})$, inference is expressed as Bayes' theorem, which relates this likelihood to both belief distributions:
\begin{equation}
	p(\boldsymbol{\theta}) 
	= \frac{L_X(\boldsymbol{\theta}) \pi(\boldsymbol{\theta})}{\int L_{\boldmath{X}}(\boldsymbol{\theta'}) \pi(\boldsymbol{\theta'}) d\boldsymbol{\theta'}_H} 
	\equiv \frac{L_{\boldmath{X}}(\boldsymbol{\theta}) \pi(\boldsymbol{\theta})}{M_{\boldmath{X}}^H},
\label{bayesthm}
\end{equation}
where the normalisation factor $M_{\boldmath{X}}^H$ is known as the marginal likelihood. In practice, the prior $\pi(\boldsymbol{\theta})$ is an educated guess, perhaps taking preceding experimental information into account, but presumed to be incomplete. As measured data becomes available, the prior probability is updated according to Bayes' theorem, and each calculated posterior probability becomes the new prior. Given enough data, this process converges to the true best-fit model regardless of error in the prior\footnote{An exception is the Dirac-delta prior, which forces the posterior to match, i.e., $p(\boldsymbol{\theta}_H) = \pi(\boldsymbol{\theta}_H) = \delta(\boldsymbol{\theta}_H - \boldsymbol{\hat{\theta}}_H)$, thereby leading to an incorrect result regardless of data. Similarly sharp prior distributions therefore converge very slowly.}.

In the analysis of typical neutrino experiments, the available data is far from sufficient to claim posterior convergence --- instead credible regions and bounds on model parameters are sought. All such quantities are expressible as posterior integrals \cite{Speagle2020}, typically written as expectation values,
\begin{equation}
    \langle f(\boldsymbol{\theta}) \rangle_p \equiv \int f(\boldsymbol{\theta}) p(\boldsymbol{\theta}) d\boldsymbol{\theta},
\end{equation}
for any integrable function $f(\boldsymbol{\theta})$.

For example, it is often desirable to obtain point estimates $\boldsymbol{\hat{\theta}}$ of a distribution, defined as single parameter sets which are representative of the sample. We can define loss functions $L(\boldsymbol{\hat{\theta}},\boldsymbol{\theta})$ and evaluate the expected loss $\langle L(\boldsymbol{\hat{\theta}},\boldsymbol{\theta})\rangle_p$. An optimal point estimate will minimize $\langle L \rangle_p$; the mean, median, and mode can be obtained by selecting loss functions $L = |\boldsymbol{\hat{\theta}} - \boldsymbol{\theta}|^2$, $|\boldsymbol{\hat{\theta}} - \boldsymbol{\theta}|$, and $I(\boldsymbol{\hat{\theta}} \neq \boldsymbol{\theta})$, respectively~\cite{Speagle2020}, where the indicator function $I$ returns $1$ if its argument is true, and $0$ otherwise.

Another use of the posterior integral is in the definition of credible regions. Here, a choice of percentile $Y$ sets a posterior threshold $p_Y$ that bounds $Y \cdot 100\%$ of the posterior from below:
\begin{equation}
    \langle I(p(\boldsymbol{\theta}_H) \geq p_Y) \rangle_p = Y.
\end{equation}

The central role of posterior integrals leads us to consider Monte Carlo methods which sample the posterior distribution, allowing for the efficient application of Bayesian inference. 

\subsection*{A \hspace{10pt} Theoretical Basis}
A typical Monte Carlo procedure simulates a target distribution $p(\boldsymbol{\theta})$ by first sampling uniformly over the domain. At each sampled point $\boldsymbol{\theta'}$, a uniform random value between zero and one is generated, and compared with $p(\boldsymbol{\theta'})$; if the random value is the lesser, the point is accepted, and otherwise thrown away. Computation of an integral $\langle X(\boldsymbol{\theta}) \rangle_p$ is then reduced to a discrete average of $X(\boldsymbol{\theta})$ values at all sample points. However, for strongly peaked distributions, the acceptance ratio during sampling rapidly approaches zero as the dimensionality increases. Importance sampling \cite{Speagle2020} and extensions thereof \cite{Gelman1998} can improve acceptance by sampling from a nonuniform proposal distribution and appropriately reweighting to reproduce the target. However, this improvement is reliant on choosing a proposal distribution sufficiently similar to the target, which is diminishingly unlikely under conditions of incomplete experimental knowledge.

Markov Chain Monte Carlo (MCMC) methods increase acceptance by considering a sequence of sampling distributions over the domain whose equilibrium distribution matches that of the target $p(\boldsymbol{\theta})$. To prevent initial sampling from having persistent effects, each sampled point $\boldsymbol{\theta}_i$ is required to depend only on the previous $\boldsymbol{\theta}{i-1}$, defining a Markov Chain, $\{\boldsymbol{\theta}_0, \boldsymbol{\theta}_1, ... \}$. In order to preserve the desired equilibrium distribution, a property known as \textit{stationarity} \cite{Brooks2011}, the transition probability $t(\boldsymbol{\theta}_i | \boldsymbol{\theta}_{i-1})$ must satisfy the \textit{detailed balance} equation \cite{Press2007},
\begin{equation}
	p(\boldsymbol{\theta}_{i-1}) t(\boldsymbol{\theta}_i | \boldsymbol{\theta}_{i-1}) = p(\boldsymbol{\theta}_i) t(\boldsymbol{\theta}_{i-1} | \boldsymbol{\theta}_i).
\end{equation}
Integrating, the posterior distribution $p(\theta_i)$ is seen to be equivalent to the superposition of transitioned states from all possible preceding iterations,
\begin{equation}
	\int t(\boldsymbol{\theta}_i | \boldsymbol{\theta}_{i-1}) p(\boldsymbol{\theta}_{i-1}) d\boldsymbol{\theta}_{i-1} = p(\boldsymbol{\theta}_i).
\end{equation}
This explicitly confirms the stationarity of the distribution from which samples are drawn as the chain converges. It further demonstrates the \textit{reversibility} property of MCMC, that the chain of samples satisfies the Markov property with time reversed \cite{Brooks2011}.

Stationarity and reversibility are manifested cleanly in the Metropolis-Hastings update procedure for MCMC \cite{Metropolis1953, Hastings1970}. A proposal distribution $q(\boldsymbol{\theta}'_i | \boldsymbol{\theta}_{i-1})$ is defined to generate candidate successor points of the Markov Chain. For candidate point $\boldsymbol{\theta}'_i$, an acceptance probability $\alpha(\boldsymbol{\theta}_{i-1},\boldsymbol{\theta}'_i)$ is constructed such that $t(\boldsymbol{\theta}'_i | \boldsymbol{\theta}_{i-1}) = \alpha(\boldsymbol{\theta}_{i-1},\boldsymbol{\theta}'_i) \cdot q(\boldsymbol{\theta}'_i | \boldsymbol{\theta}_{i-1})$ satisfies the detailed balance equation,
\begin{equation}
	\alpha(\boldsymbol{\theta}_{i-1},\boldsymbol{\theta}'_i) = \text{min}\left[ 1, \frac{p(\boldsymbol{\theta}'_i) q(\boldsymbol{\theta}_{i-1} | \boldsymbol{\theta}'_i)}{p(\boldsymbol{\theta}_{i-1}) q(\boldsymbol{\theta}'_i | \boldsymbol{\theta}_{i-1})} \right].
\end{equation}
When the proposal function $q$ is symmetric between $\boldsymbol{\theta}_{i-1}$ and $\boldsymbol{\theta}'_i$, the right-hand ratio cancels and simplifies this computation significantly. The choice of proposal distribution must be tuned to the application: wider distributions will more efficiently hop between disconnected regions of a distribution, but may result in lower acceptance rates. Regardless of this selection, so long as the detailed balance equation is satisfied, reversibility of the Markov chain may be proven by evaluating the transition probability in both directions between two general points in the parameter space. The central limit theorem for Markov chains \cite{Brooks2011} then holds, and together with technical conditions such as Harris recurrence (that the chain must be able to eventually return to every point in its parameter space), proves that the chain will converge to the equilibrium distribution regardless of initial sampling distribution. Naturally, other update procedures exist; in particular Gibbs and Hamiltonian MCMC are frequently used, as are adaptive MCMC approaches where the proposal distribution is adjusted in response to the acceptance rate during runtime. We found these update procedures to exceed our needs.

We chose to retain flat priors on model parameters throughout this work, with the notable exceptions of multivariate Gaussian priors applied to light and heavy NMEs. For neutrino mass parameters in particular, the comparative advantages of flat and logarithmic priors are considered throughout the literature, and in a previous work the authors considered the impact of least-informative priors on inference over the $0\nu\beta\beta$ decay parameter space \cite{Deppisch2021}.

\subsection*{B \hspace{10pt} Convergence Diagnostics}
It is not generally possible to confirm from purely \textit{a priori} (i.e., physical) considerations whether an MCMC has converged. Particularly for multimodal distributions in high dimensions, chains can become temporarily trapped in local optima, and a substantial runtime may be required to appropriately sample all regions. Even for unimodal distributions over a single parameter, single chains of finite length can have misleading convergence properties which are not intrinsically avoidable. 

Gelman and Rubin \cite{Gelman1992} propose a two-part strategy for improving and assessing convergence. First, in analogy to iterative optimisation problems, where multiple optima must be identified by multiple independent initialisations, multiple MCMC sequences are suggested to improve convergence. The initialisations could in principle be drawn from a broadened approximate to the posterior distribution, although we chose in this work to draw them flatly across the parameter space, and found no ill effect given a sufficient burn-in period.

Second, during MCMC runtime, convergence is monitored by evaluating the rate of mixing between independent chains. Any distinguishability between chains is taken to indicate incomplete or failed convergence, for by the stationarity property, each chain may be expected to converge to the same posterior distribution regardless of initialisation. The resultant \textit{potential scale reduction} $\sqrt{R}$ represents the first and foremost quantitative metric of MCMC convergence, whereas all previous metrices relied on qualitative, graphical methods.

Following the notation of \cite{Vehtari2020}, $W$ denotes the average variance within an independent chain, and $B$ the variance between the means of the chains. Letting each of $M$ chains $\boldsymbol{\theta}^m$ contain $N$ samples $\boldsymbol{\theta}^m_n$,
\begin{equation}
    W = \frac{1}{M}\sum_{m=1}^M V_n(\boldsymbol{\theta}^m), \hspace{40pt}
    B = V_m\left(\frac{1}{N}\sum_{n=1}^N \boldsymbol{\theta}^m_n\right),
\end{equation}
where $V_x$ denotes the variance taken over index $x$.
The variance of the posterior distribution is then well-approximated by a weighted average of the within- and between-chain variances:
\begin{equation}
    \sigma^2 \approx \frac{N-1}{N}W + \frac{1}{N}B.
\end{equation}
Initially, and whenever chains are poorly mixed, one can expect $B \gg W$ and therefore a significant mismatch between the overall variance estimate $\sigma^2$ and the variance of any individual chain. However once all chains have converged according to the stationarity property, $W$ must approach the value of $\sigma^2$. As a result, the potential scale reduction is defined,
\begin{equation}
    \sqrt{R} \equiv \sqrt{\frac{\sigma^2}{W}}.
\end{equation}
A $\sqrt{R}$ value near to 1 provides strong evidence that all chains have a sampling distribution which accurately reflects the posterior distribution. One flaw in this metric is that $\sqrt{R} = 1$ can occur for chains which have fully explored the parameter space but not yet individually converged, if the drift directions of those chains cancel each other out. Ref.~\cite{Gelman1992} avoids this heuristically by including $N$ initial ``burn-in'' samples in each chain, which are not included in any posterior estimate. Ref.~\cite{Vehtari2020} takes the further direct approach of bisecting each of the $M$ chains and evaluating $\sqrt{R}$ for the resultant $2M$ segments.

For all simulations shown in this work, we made use of the general-purpose MCMC library implemented in C++11 in \cite{Deppisch2021}. The Metropolis-Hastings update scheme was favoured, and equipped to handle parameter spaces of any dimensionality. Pseudorandom numbers were generated according to STL random device MT19937, the Mersenne Twister algorithm with prime-number period $2^{19937}-1$. All visualisations shown in this work were performed using Matplotlib.

\bibliographystyle{JHEP}
\bibliography{bib}

\end{document}